\documentclass[smallextended]{svjour3}       % onecolumn (second format)
\smartqed  % flush right qed marks, e.g. at end of proof
\usepackage{graphicx,color,amsmath,amssymb,epsfig}

\newcommand{\du}{\mbox{d}u}
\newcommand{\dv}{\mbox{d}v}
\newcommand{\dt}{\mbox{d}t}
\numberwithin{equation}{section}
\newcommand{\picturesAB}[4]{
\centerline{
\hskip #4
\raise #3 \hbox{\raise 0.9mm \hbox{(a)}}
\hskip -5mm
\includegraphics[height=#3]{#1}
\hskip 2mm
\raise #3 \hbox{\raise 0.9mm \hbox{(b)}}
\hskip -5mm
\includegraphics[height=#3]{#2}
}}
\newcommand{\picturesCD}[4]{
\centerline{
\hskip #4
\raise #3 \hbox{\raise 0.9mm \hbox{(c)}}
\hskip -5mm
\includegraphics[height=#3]{#1}
\hskip 2mm
\raise #3 \hbox{\raise 0.9mm \hbox{(d)}}
\hskip -5mm
\includegraphics[height=#3]{#2}
}}
\newcommand{\picturesnoAB}[4]{
\centerline{
\hskip #4
\includegraphics[height=#3]{#1}
\hskip 2mm
\includegraphics[height=#3]{#2}
}}
\newcommand{\picturesABal}[5]{
\centerline{
\hskip #4
\raise #3 \hbox{\raise 0.9mm \hbox{(a)}}
\hskip -5mm
\includegraphics[height=#3]{#1}
\hskip #5
\raise #3 \hbox{\raise 0.9mm \hbox{(b)}}
\hskip -5mm
\includegraphics[height=#3]{#2}
}}
\newcommand{\picturesABalt}[5]{
\centerline{
\hskip #4
\raise #3 \hbox{\raise 3mm \hbox{(a)}}
\hskip -5mm
\includegraphics[height=#3]{#1}
\hskip #5
\raise #3 \hbox{\raise 3mm \hbox{(b)}}
\hskip -5mm
\includegraphics[height=#3]{#2}
}}

\journalname{Submitted to Bulletin of Mathematical Biology}

\begin{document}

\title{Stochastic Turing patterns: analysis of compartment-based
approaches}

\author{Yang Cao \and Radek Erban}

\institute{Yang Cao \at
Department of Computer Science, Virginia Tech,  
Blacksburg, VA 24061, USA \\
\email{ycao@vt.edu}
\and
Radek Erban \at
Mathematical Institute, University of Oxford \\
Radcliffe Observatory Quarter, Woodstock Road, 
Oxford, OX2 6GG, United Kingdom \\
\email{erban@maths.ox.ac.uk}
}

\date{Preprint version: \today}

\maketitle

\begin{abstract}
Turing patterns can be observed in reaction-diffusion 
systems where chemical species have different diffusion
constants. In recent years, several studies investigated the
effects of noise on Turing patterns and showed that the parameter
regimes, for which stochastic Turing patterns are observed, can
be larger than the parameter regimes predicted by deterministic
models, which are written in terms of partial differential equations
for species concentrations. A common stochastic reaction-diffusion 
approach is written in terms of compartment-based (lattice-based) 
models, where the domain of interest is divided into artificial
compartments and the number of molecules in each compartment is 
simulated. In this paper, the dependence of stochastic Turing patterns 
on the compartment size is investigated. It has previously been shown 
(for relatively simpler systems) that a modeller should not choose
compartment sizes which are too small or too large, and that the optimal 
compartment size depends on 
the diffusion constant. Taking these results into account, we propose 
and study a compartment-based model of Turing patterns where
each chemical species is described using a different set of compartments.
It is shown that the parameter regions where spatial patterns form
are different from the regions obtained by classical deterministic
PDE-based models, but they are also different from the results
obtained for the stochastic reaction-diffusion models which use
a single set of compartments for all chemical species. In particular,
it is argued that some previously reported results on the effect of noise
on Turing patterns in biological systems need to be reinterpreted.
\end{abstract}


\section{Introduction}

In his pioneering work, Alan Turing \cite{Turing:1952:CBM} showed that
stable spatial patterns can develop in reaction-diffusion systems which 
include chemical species (morphogens) with different diffusion constants. 
Considering a system of two chemical species with concentrations $u(x,t)$
and $v(x,t)$ in one-dimensional interval $x \in [0,L]$, the
underlying deterministic model of Turing patterns can be written 
as a system of two reaction-diffusion partial differential equations (PDEs)
\begin{eqnarray}
\frac{\partial u}{\partial t}
&=&
D_u
\frac{\partial^2 u}{\partial x^2}
+
f_1(u,v),
\label{uPDE}
\\
\frac{\partial v}{\partial t}
&=&
D_v
\frac{\partial^2 v}{\partial x^2}
+
f_2(u,v),
\label{vPDE}
\end{eqnarray}
where $D_u$ and $D_v$ are diffusion constants of morphogens $u$ and $v$,
respectively, and $f_1(u,v)$ and $f_2(u,v)$ describe chemical reactions.
Then the standard analysis proceeds as follows
\cite{Murray:2002:MB,Satnoianu:2000:TIG}:
a homogeneous steady state $u(x,t) \equiv u_s$, $v(x,t) \equiv v_s$
is found by solving $f_1(u_s,v_s) = 0$ and $f_2(u_s,v_s)=0$. It is
shown that the homogenous steady state is stable when $D_u = D_v$,
and conditions on $f_1$, $f_2$, $D_u$ and $D_v$ are obtained which
guarantee that the homogeneous steady state will become unstable for 
$D_u \ne D_v$. Then Turing patterns are observed at the steady
state.

The above argument was extensively analysed in the mathematical 
biology literature and conditions for Turing patterns have been
determined \cite{Murray:2002:MB,Satnoianu:2000:TIG}. Experimental 
studies with chemical systems (chlorite-iodide-malonic acid
reaction) demonstrated Turing type patterns
\cite{Kepper:1991:TCP,Quyang:1991:TUS}. There has also
been experimental evidence that a simple Turing patterning mechanism
can appear in developmental biology, for example, in the regulation of
hair follicle patterning in developing murine skin \cite{Sick:2006:WDD}.
One of the criticism of Turing patters is their lack of 
robustness \cite{Maini:2012:TMB}.
The PDE system (\ref{uPDE})--(\ref{vPDE}) can have several stable
non-homogeneous solutions which the system can achieve with relatively
small perturbations to the initial condition. Considering PDEs in
a suitably growing domain, one can obtain an additional constraint on 
the system which restricts the set of accessible patterns, increasing 
the robustness of pattern generation with respect to the initial 
conditions \cite{Crampin:1999:RDG,Barrass:2006:MTM}. However, to 
assess the sensitivity of patterns with respect to fluctuations,
stochastic models have to be considered \cite{Maini:2012:TMB,Black:2012:SFE}.

One of the most common approaches to stochastic reaction-diffusion
modelling is formulated in the compartment-based (lattice-based)
framework \cite{Erban:2007:PGS}. In the one-dimensional
setting, the compartment-based analogue of the PDE model
(\ref{uPDE})--(\ref{vPDE}) can be formulated as follows:
The computational domain $[0,L]$ is divided into $K$ compartments 
of length $h = L/K$. We denote the number of molecules of chemical 
species $U$ (resp. $V$) in the $i$-th compartment $((i-1)h,ih)$ by 
$U_i$ (resp. $V_i$), $i=1,2,\dots,K$. Then the diffusion of $U$
and $V$ is described by the following chains of 
``chemical reactions" \cite{Erban:2007:PGS}:
\begin{equation}
U_1
\;
\mathop{\stackrel{\displaystyle\longrightarrow}\longleftarrow}^{d_u}_{d_u}
\;
U_2
\;
\mathop{\stackrel{\displaystyle\longrightarrow}\longleftarrow}^{d_u}_{d_u}
\;
U_3
\;
\mathop{\stackrel{\displaystyle\longrightarrow}\longleftarrow}^{d_u}_{d_u}
\;
\dots
\;
\mathop{\stackrel{\displaystyle\longrightarrow}\longleftarrow}^{d_u}_{d_u}
\;
U_K,
\label{diffGillU}
\end{equation}
\begin{equation}
V_1
\;
\mathop{\stackrel{\displaystyle\longrightarrow}\longleftarrow}^{d_v}_{d_v}
\;
V_2
\;
\mathop{\stackrel{\displaystyle\longrightarrow}\longleftarrow}^{d_v}_{d_v}
\;
V_3
\;
\mathop{\stackrel{\displaystyle\longrightarrow}\longleftarrow}^{d_v}_{d_v}
\;
\dots
\;
\mathop{\stackrel{\displaystyle\longrightarrow}\longleftarrow}^{d_v}_{d_v}
\;
V_K
\label{diffGillV}
\end{equation}
where
\begin{equation}
d_u = \frac{D_u}{h^2}
\qquad \mbox{and} \qquad
d_v = \frac{D_v}{h^2}.
\label{dudvformula}
\end{equation}
Reactions are localized to each compartment. For example,
considering the commonly studied Schnakenberg reaction system
\cite{Schnakenberg:1979:SCR}, chemical reactions in the
$i$-th compartment are described by \cite{Qiao:2006:SDS}:
\begin{equation}
{\mbox{ \raise 0.851 mm \hbox{$\emptyset$}}}
\;
\mathop{\stackrel{\displaystyle\longrightarrow}\longleftarrow}^{k_1}_{k_2}
\;
{\mbox{\raise 0.851 mm\hbox{$U_i$,}}}
\qquad\qquad
\mbox{ \raise 1mm \hbox{%
 $\emptyset \;
 \displaystyle\mathop{\displaystyle\longrightarrow}^{k_3}\; V_i$,}}
\qquad\qquad
\mbox{ \raise 1mm \hbox{%
$2 U_i + V_i 
\;\displaystyle\mathop{\displaystyle\longrightarrow}^{k_4}\; 3 U_i$.}}
\label{schnak}
\end{equation}
The above formulation (\ref{diffGillU}), (\ref{diffGillV}) and
(\ref{schnak}) describes the stochastic reaction-diffusion model
as a system of ($8K-4$) chemical reactions: we have ($K-1$) diffusive
jumps of $U$ molecules to the left (resp. right),
($K-1$) diffusive
jumps of $V$ molecules to the left (resp. right),
and $4 K$ reactions (\ref{schnak}). This system can be simulated
using the Gillespie algorithm \cite{Gillespie:1977:ESS},
or its equivalent formulations \cite{Cao:2004:EFS,Gibson:2000:EES}.
In Figure \ref{figure1}, we present an illustrative simulation
of the reaction-diffusion system (\ref{diffGillU}), 
(\ref{diffGillV}) and (\ref{schnak}). We clearly see that Turing
patterns can be observed for the chosen set of dimensionless
parameters:
\begin{figure}
\picturesAB{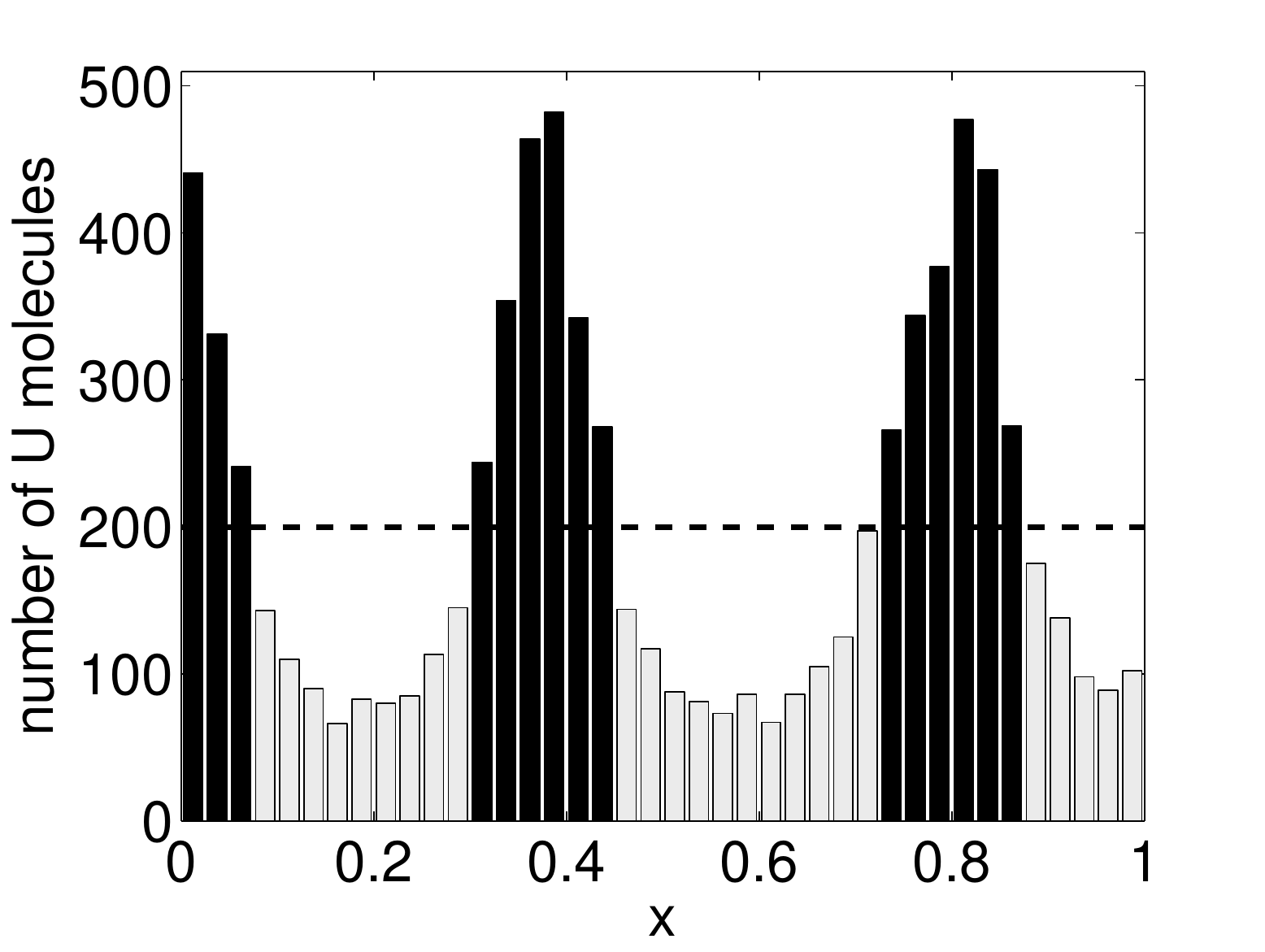}{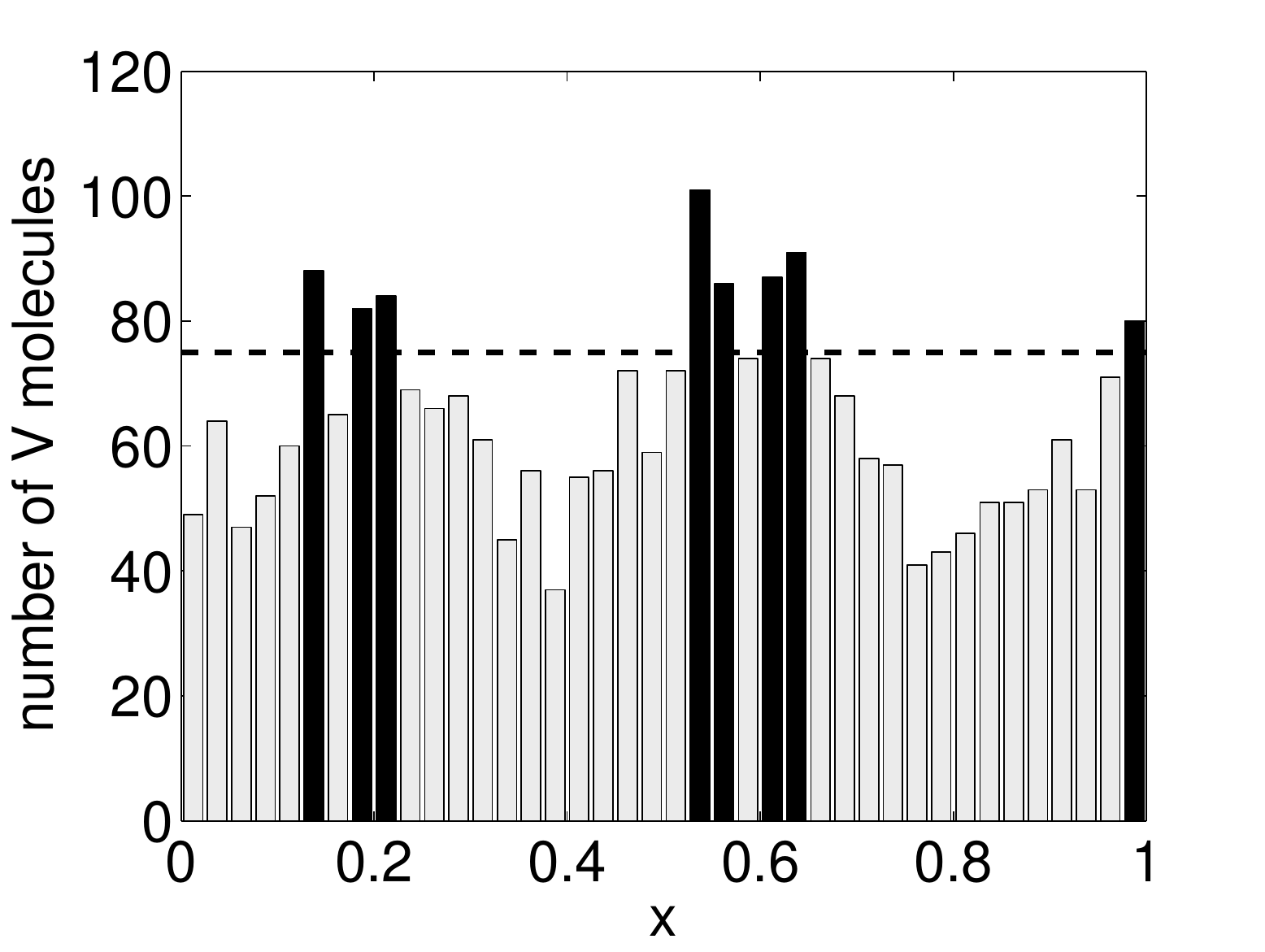}{4.5cm}{5mm}
\caption{{\it Turing patterns for the stochastic reaction-diffusion system 
$(\ref{diffGillU})$, $(\ref{diffGillV})$ and $(\ref{schnak})$. 
{\rm (a)} Numbers of molecules of chemical species $U$ in 
each compartment at time $18$;
{\rm (b)} the same plot for chemical species $V$. The initial
condition was the homogeneous steady state $U_{st} = 200$ and
$V_{st} = 75$ for the parameters given in the text. The values
of $U_{st}$ and $V_{st}$ are denoted by dashed lines. 
Adapted from {\rm \cite{Erban:2007:PGS}} with permission.}}
\label{figure1}
\end{figure}
$k_1 = 4 \times 10^3$, $k_2 = 2$, $k_3 = 1.2 \times 10^3$,
$k_4 = 6.25 \times 10^{-8}$, $D_u = 10^{-3}$ and $D_v = 10^{-1}$. 
Compartment values above (resp. below) the homogeneous steady state
values $U_{st} = 200$ and $V_{st} = 75$ are coloured black (resp. 
light gray) to visualize stochastic Turing patterns. Let us note that 
the rate constants $k_1$ and $k_3$ are production rates per unit
of area. The stochastic model uses the production rates per 
one compartment which are given as $k_1 h$ and $k_3 h$, respectively. 
More details of this stochastic simulation are given in 
Section \ref{sec2} where we introduce the corresponding 
propensity functions (\ref{firstfouralpha})--(\ref{lastfouralpha}).

The compartment-based approach has been used for both theoretical
analysis and computational modelling \cite{Scott:2011:AIN,Hattne:2005:SRD}. 
The regions where stochastic Turing patterns can be expected were calculated 
using the linear noise analysis
\cite{Biancalani:2010:STP,McKane:2013:SPF,Butler:2011:FTP}.
These studies were also generalized to growing domains  
\cite{Woolley:2011:PSM,Woolley:2011:SRD}, to stochastic
reaction-diffusion models with delays \cite{Woolley:2012:EIS},
to non-local trimolecular reactions \cite{Biancalani:2011:SWB}
and to stochastic Turing patterns on a network \cite{Asslani:2012:STP}.
Compartment-based software packages were developed \cite{Hattne:2005:SRD}
and applied to modelling biological systems \cite{Fange:2006:NMP}.
Computational approaches were also generalized to non-regular
compartments (lattices) and complex geometries
\cite{Engblom:2009:SSR,Isaacson:2006:IDC}. Stochastic 
simulations of Turing patterns 
\cite{Twomey:2007:SMR,Fu:2008:SST,Hori:2012:NSP}
and excitable media \cite{Vigelius:2012:SSP}
were also presented in the literature.
However, these theoretical and computational studies use the same 
discretization for each chemical species. In this paper, we will 
demonstrate that, in the case of Turing patterns, this simplifying 
assumption can undesirably bias the obtained
theoretical and computational results.

\begin{figure}
\picturesABalt{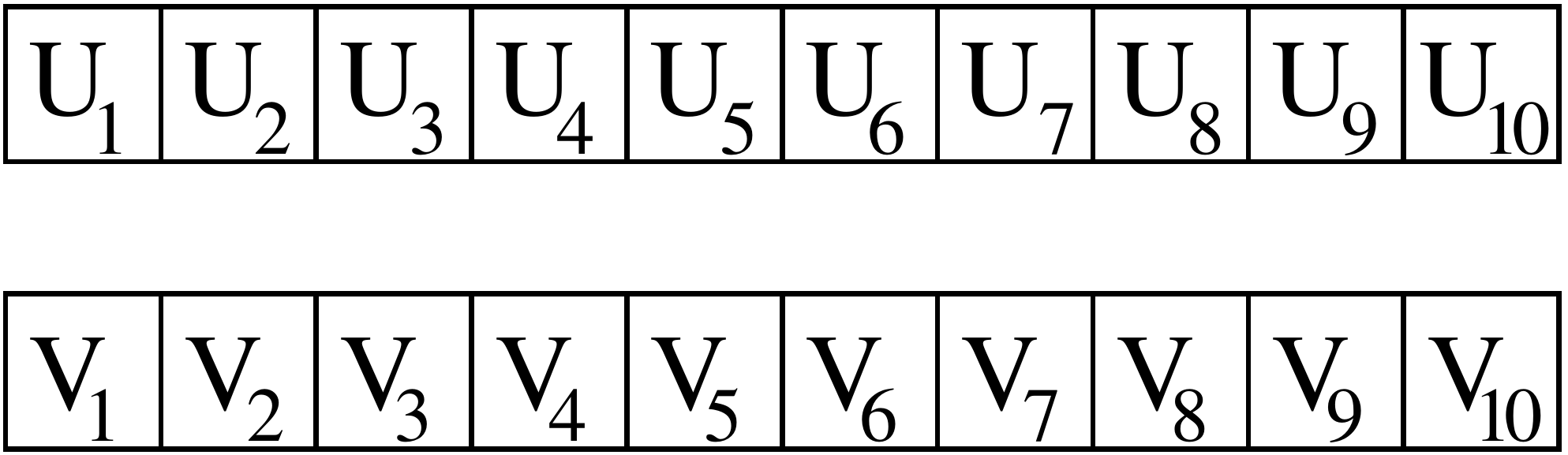}{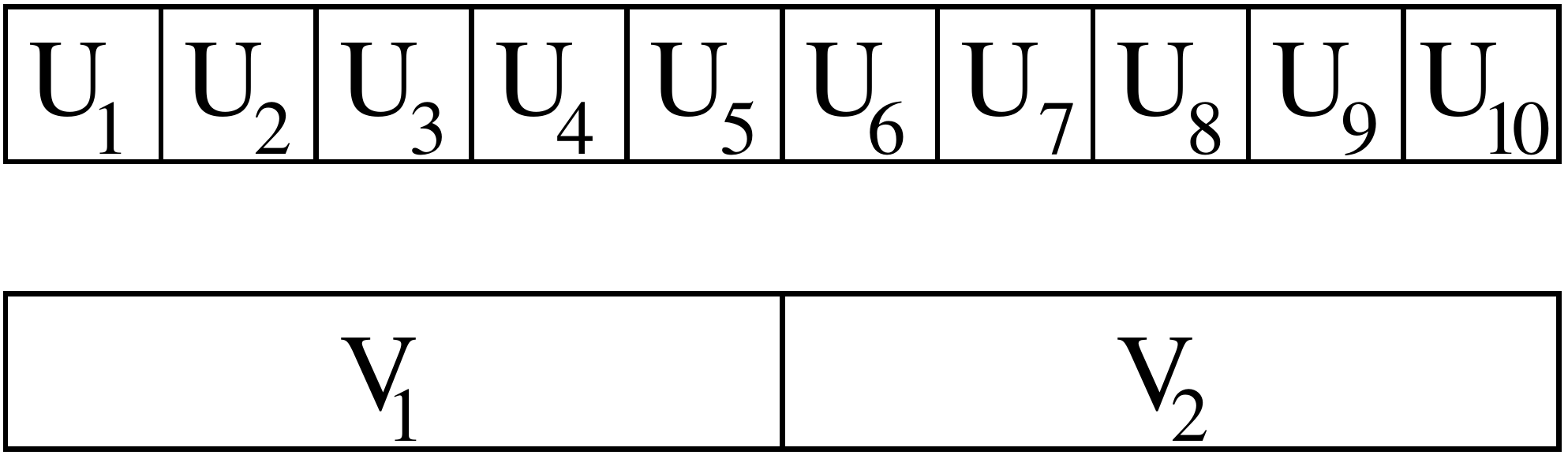}{1.4cm}{2mm}{12mm}
\caption{
(a) {\it Schematic of the uniform discretization.}
(b) {\it Schematic of different meshes used for $U$ and $V$
where $\gamma$ defined by $(\ref{definitionhuhv})$ is equal to 5.}}
\label{figure2}
\end{figure}

One of the assumption of the compartment-based modelling is that
compartments are small enough so that they can 
be assumed well-mixed. In particular,
the relative size of diffusion and reaction constants determine
the appropriate size of the compartment 
\cite{Erban:2009:SMR,Isaacson:2009:RME,Hellander:2012:RME}.
It can be shown that there exists a limitation on the compartment
size from below whenever the reaction-diffusion system
includes a bimolecular reaction 
\cite{Erban:2009:SMR,Isaacson:2009:RME,Hellander:2012:RME}.
There are also bounds on the compartment size from above
\cite{Kang:2012:NMC,Hu:2013:SAR}, again the diffusion constant
plays an important role in these estimates. In the case of Turing
patterns, we have chemical species with different diffusion
constants. For example, in the illustrative
simulation in Figure \ref{figure1}, we have
$D_v/D_u = 100$, i.e. the diffusion constant of $V$ is
100-times larger than the diffusion constant of $U$.
However, we used the same discretization for both $U$ and $V$
which is schematically denoted in Figure \ref{figure2}(a).
If we take into account that $V$ diffuses much faster, then
one could also consider the discretization
in Figure \ref{figure2}(b) where one compartment in the $V$ variable
corresponds to several compartments in the $U$ variable. In this
paper, we will study differences between discretizations 
in Figure \ref{figure2}(a) and Figure \ref{figure2}(b).
We will show that these discretizations lead to different
parameter regimes for stochastic Turing patterns. 

The paper is
organized as follows. In Section \ref{sec2} we introduce and
analyse a simple test problem which will be used to illustrate our
results. It will be based on the above model 
(\ref{diffGillU}), (\ref{diffGillV}) and (\ref{schnak}).
In Section \ref{sec3} we analyse both types of discretizations,
considering a simple two-compartment discretization in $U$.
Illustrative numerical results are presented in Section \ref{sec4}.
We conclude this paper with the discussion of our
results in Section \ref{sec5}.
 
\section{Deterministic and stochastic models of an illustrative
reaction-diffusion system}
\label{sec2}

We will consider a simple one-dimensional Schnakenberg model 
(\ref{schnak}) where the reaction rate constants are given
by \cite{Qiao:2006:SDS}
\begin{equation}
k_1 = \omega, \qquad
k_2 = 2, \qquad
k_3 = 3\omega, \qquad
k_4 = \frac{1}{\omega^2}
\label{omegascaling}
\end{equation}
and $\omega$ is a scale factor. We used $\omega = 4 \times 10^3$
in the illustrative simulation in Figure \ref{figure1}.
When there is no diffusion involved, the dynamics of this 
system can be represented as the system of reaction rate ordinary
differential equations (ODEs)
\begin{eqnarray*} 
\frac{\du}{\dt} & = & k_1 - k_2 u + k_4 u^2 v, \\
\frac{\dv}{\dt} & = & k_3 - k_4 u^2 v, 
\end{eqnarray*} 
which has a unique stable steady state at $u_s = 2 \omega$ and 
$v_s = 3\omega/4$. When we consider diffusion, the 
reaction-diffusion PDEs (\ref{uPDE})--(\ref{vPDE}) are given by
\begin{eqnarray}
\frac{\partial u}{\partial t}
&=&
D_u
\frac{\partial^2 u}{\partial x^2}
+
k_1 - k_2 u + k_4 u^2 v,
\label{uPDEschnak}
\\
\frac{\partial v}{\partial t}
&=&
D_v
\frac{\partial^2 v}{\partial x^2}
+
k_3 - k_4 u^2 v.
\label{vPDEschnak}
\end{eqnarray}
We are implicitly assuming homogeneous Neumann boundary conditions
(zero-flux) in the whole paper, but both the PDE model 
(\ref{uPDEschnak})--(\ref{vPDEschnak}) and its stochastic
counterparts could also be generalized to different types of boundary
conditions \cite{Erban:2007:RBC}. Using standard analysis of
Turing instabilities \cite{Qiao:2006:SDS,Murray:2002:MB},
one can show that the Turing patterns are obtained for
$D_v > 39.6 D_u$ for the parameter values (\ref{omegascaling}).
This condition is independent of $\omega.$ The illustrative
simulation in Figure \ref{figure1} was computed for $D_v/D_u = 100$,
i.e. the condition for (deterministic, mean-field) Turing patterns 
was satisfied.

When we are concerned with the stochastic effects, the reaction-diffusion 
system can be simulated by the Gillespie stochastic simulation algorithm 
with the one-dimensional computational domain $[0,L]$ discretized. 
Considering uniform discretization
in Figure \ref{figure2}(a), the stochastic model is given as
a set of ``chemical reactions"
(\ref{diffGillU}), (\ref{diffGillV}) and (\ref{schnak}).
Denoting the compartment length by $h$, we have the following propensity 
functions in the $i$-th compartment \cite{Gillespie:1977:ESS,Qiao:2006:SDS}: 
\begin{equation}
\alpha_1 = k_1 h, \quad
\alpha_2 = k_2 U_i,  \quad
\alpha_3 = k_3 h, \quad
\alpha_4 = \frac{k_4}{h^2} U_i (U_i - 1) V_i, 
\label{firstfouralpha}
\end{equation}
\begin{equation}
\alpha_5 = \alpha_6 = d_u U_i, \qquad\qquad
\alpha_7 = \alpha_8 = d_v V_i, 
\label{lastfouralpha}
\end{equation} 
where $d_u$ and $d_v$ are given by (\ref{dudvformula}).
The first four propensities (\ref{firstfouralpha})
are for the four chemical reactions in \eqref{schnak}.
The propensities (\ref{lastfouralpha}) are for the diffusive 
jumps (left and right) for $U$ (indices 5 and 6) and 
$V$ (indices 7 and 8) which correspond to
(\ref{diffGillU}) and (\ref{diffGillV}), respectively.
In the illustrative simulation in Figure \ref{figure1},
we divided interval $[0,1]$ into $K=40$ compartments,
i.e. $h=1/40=0.025$. In particular, the production
rate of $U$ molecules in one compartment was
equal to $\alpha_1 = k_1 h = \omega h = 100$. The homogeneous
steady state in compartments corresponded to values
$U_{st} = u_s h = 2 \omega h = 200$ and
$V_{st} = v_s h = 3 h \omega /4 = 75.$

\subsection{Formulation of the generalized comparment-based model}

The compartmentalization in Figure \ref{figure2}(b) generalizes
(\ref{diffGillU}) and (\ref{diffGillV}) to the case where different
discretizations are used for $U$ and $V$. We will denote by
$K_u$ (resp. $K_v$) the number of compartments in the $U$ 
(resp. $V$) variable. We define the compartment lengths by
\begin{equation}
h_u
= 
\frac{L}{K_u},
\qquad
h_v
=
\frac{L}{K_v},
\qquad
\mbox{and}
\qquad
\gamma = \frac{K_u}{K_v} = \frac{h_v}{h_u},
\label{definitionhuhv}
\end{equation}
where $\gamma$ is the ratio of compartment sizes in the $V$ and $U$ variable.
In what follows, we will consider that $\gamma$ is an integer. For example,
a schematic diagram in Figure \ref{figure2}(b) used
$\gamma=5$. Then the diffusion model is formulated
as follows 
\begin{equation}
U_1
\;
\mathop{\stackrel{\displaystyle\longrightarrow}\longleftarrow}^{d_u}_{d_u}
\;
U_2
\;
\mathop{\stackrel{\displaystyle\longrightarrow}\longleftarrow}^{d_u}_{d_u}
\;
U_3
\;
\mathop{\stackrel{\displaystyle\longrightarrow}\longleftarrow}^{d_u}_{d_u}
\;
\dots
\;
\mathop{\stackrel{\displaystyle\longrightarrow}\longleftarrow}^{d_u}_{d_u}
\;
U_{K_u},
\label{diffGillUm}
\end{equation}
\begin{equation}
V_1
\;
\mathop{\stackrel{\displaystyle\longrightarrow}\longleftarrow}^{d_v}_{d_v}
\;
V_2
\;
\mathop{\stackrel{\displaystyle\longrightarrow}\longleftarrow}^{d_v}_{d_v}
\;
V_3
\;
\mathop{\stackrel{\displaystyle\longrightarrow}\longleftarrow}^{d_v}_{d_v}
\;
\dots
\;
\mathop{\stackrel{\displaystyle\longrightarrow}\longleftarrow}^{d_v}_{d_v}
\;
V_{K_v},
\label{diffGillVm}
\end{equation}
where
\begin{equation}
d_u = \frac{D_u}{h_u^2},
\qquad\qquad
d_v = \frac{D_v}{h_v^2} = \frac{D_v}{D_u \gamma^2} \, d_u.
\label{dudvformulam}
\end{equation}
In the standard comparment-based model (\ref{diffGillU}) 
and (\ref{diffGillV}), we have $\gamma=1$. One option to choose
$\gamma$ in the generalized model (\ref{diffGillUm}) 
and (\ref{diffGillVm}) is to ensure that $d_u = d_v$ which
implies
\begin{equation}
\gamma = \sqrt{ \frac{D_v}{D_u} }.
\label{gammaformula}
\end{equation}
Then the jump rates $d_u$ and $d_v$ from the corresponding compartments
are equal for molecules of $U$ and $V$. However,
we will not restrict to the case (\ref{gammaformula}) and consider
general choices of $\gamma$ in this paper. 
The generalization of the first three
propensities in (\ref{firstfouralpha}) is straightforward. 
Propensities $\alpha_1$ and $\alpha_2$ in (\ref{firstfouralpha}) correspond 
to chemical species $U$ and we have the following propensities in the $i$-th 
compartment, $i=1,2,\dots,K_u$: $\alpha_1 = k_1 h_u$ and $\alpha_2 = k_2 U_i$.
The propensity $\alpha_3$ in (\ref{firstfouralpha}) is considered
in the $j$-th compartment corresponding to the $V$ species, i.e.
in the compartment $\big((j-1)h_v,j h_v \big)$. It is given
as $\alpha_3 = k_3 h_v$. To generalize $\alpha_4$, we have to consider 
the occurrences of the trimolecular reaction
$$
2 U + V \;\displaystyle\mathop{\displaystyle\longrightarrow}^{k_4}\; 3 U
$$
in every small compartment in discretization of the $U$ variable.
In the $i$-th compartment, the propensity function $\alpha_4$ is:
\begin{equation}
\alpha_4 = \frac{k_4}{h_u^2} U_i (U_i - 1) \frac{V_j}{\gamma}, 
\label{generalizationofalpha4}
\end{equation}
where $V_j$ corresponds to the $j$-th compartment in the $V$ variable
to which the $i$-th compartment belongs, i.e.
$$
\big((i-1)h_u,i h_u \big) \subset \big((j-1)h_v,j h_v \big).
$$
The main idea of the compartment-based model is that the 
molecules of $V$ are considered to be well-mixed in the compartments
of the size $h_v$. Thus the propensity function (\ref{generalizationofalpha4})
correctly generalizes the propensity of trimolecular reaction $\alpha_4$
in the smaller compartment of length $h_u$.

\begin{figure}
\picturesAB{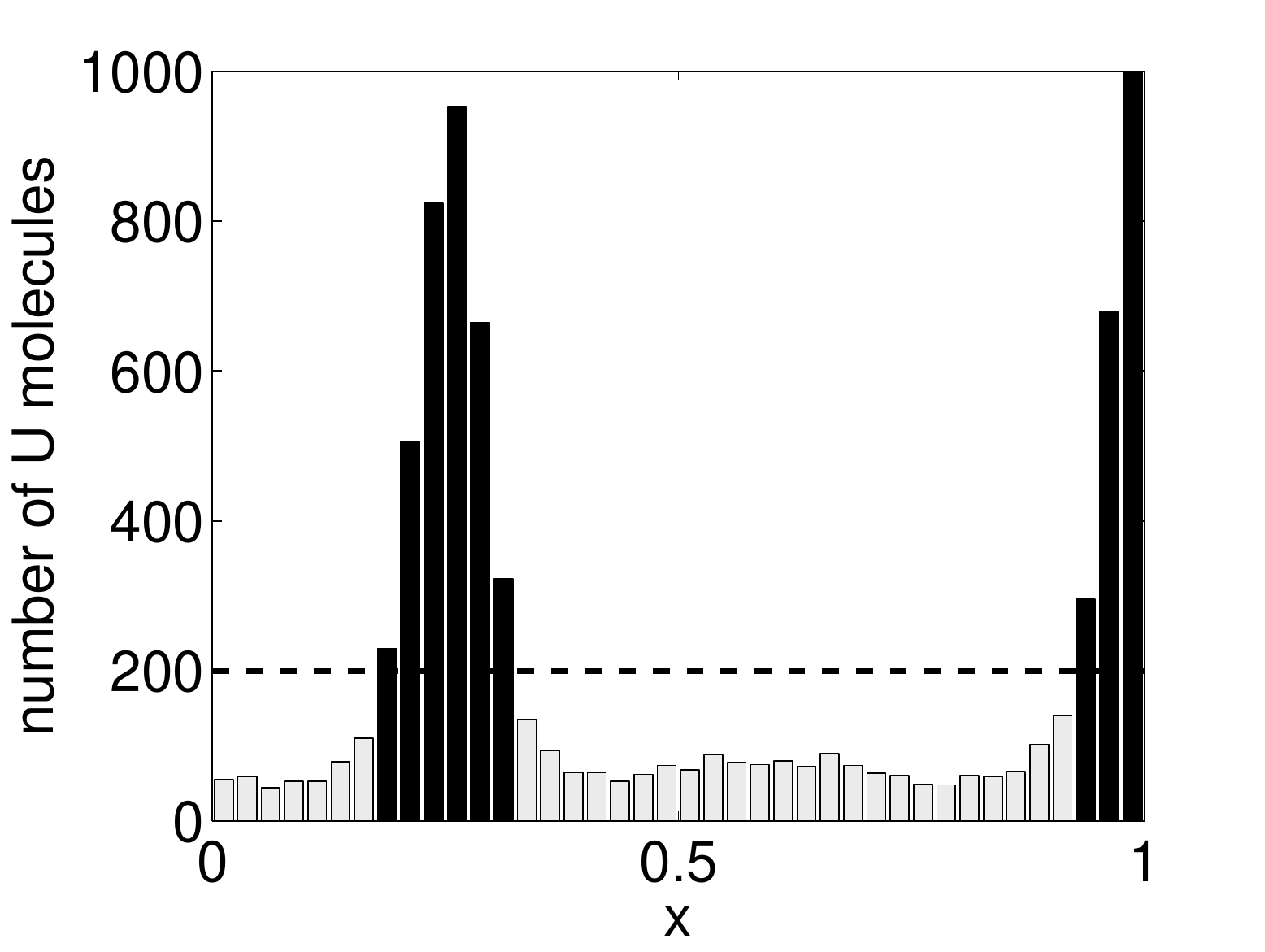}{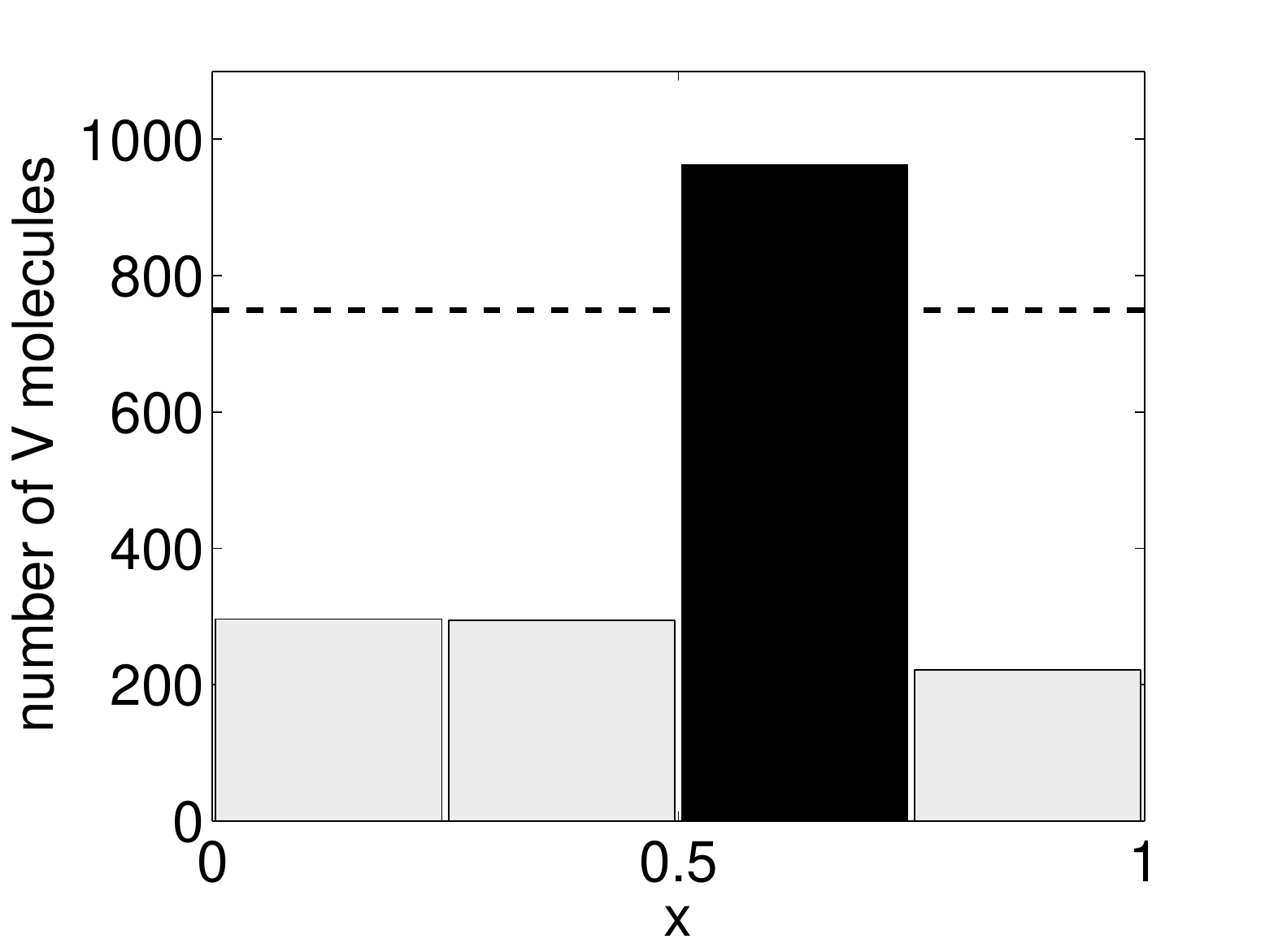}{4.5cm}{5mm}
\caption{{\it Turing patterns computed by the generalized 
compartment-based model $(\ref{diffGillUm})$--$(\ref{generalizationofalpha4})$.
{\rm (a)} Numbers of molecules of chemical species $U$ in 
each compartment at time $18$;
{\rm (b)} the same plot for chemical species $V$. The initial
condition was the homogeneous steady state $U_{st} = 200$ and
$V_{st} = 750$ for the parameters given in the text. The values
of $U_{st}$ and $V_{st}$ are denoted by dashed lines.}}
\label{figure3}
\end{figure}

In Figure \ref{figure3}, we present an illustrative simulation
of the generalized compar\-tment-based model
(\ref{diffGillUm})--(\ref{generalizationofalpha4}).
We use the same parameters as in Figure \ref{figure1} to enable direct
comparisons, i.e. $k_1$, $k_2$, $k_3$, $k_4$ are given
by (\ref{omegascaling}) where the scale factor $\omega = 4 \times 10^3$.
We use (\ref{gammaformula}) to select the value of $\gamma$.
Since $D_u = 10^{-3}$ and $D_v = 10^{-1}$, the formula
(\ref{gammaformula}) implies $\gamma = 10$. We use the same number
of compartments for $U$ variable as in Figure \ref{figure1}: $K_u = 40$. 
Using $\gamma = 10$, we obtain that $V$ is discretized into $K_v = 4$
compartments. In Figure \ref{figure3}, we see that the Turing
pattern can still be clearly observed. As in Figure \ref{figure1},
compartment values above (resp. below) the homogeneous steady state
values $U_{st} = 200$ and $V_{st} = 75 \gamma = 750$ are coloured black 
(resp. light gray) to visualize stochastic Turing patterns. 

Since the compartments in $V$ variable are 10-times larger in
Figure \ref{figure3}(b) then in Figure \ref{figure1}(b), it is not
suprising that the numbers of molecules of $V$ (per compartment) 
increased by the factor of 10. However, we can also notice that 
the numbers of molecules of $U$ per compartment quantitatively 
differ in Figure \ref{figure1}(a) and Figure \ref{figure3}(a) 
(black peaks are twice taller). An open question is to quantify
these differences. In this paper, we will study even more fundamental
issue: we will see that we can find parameter regimes where
the generalized compartment-based model exhibits Turing patterns,
while the original discretization does not.

The generalized compartment-based model (\ref{diffGillUm}) 
and (\ref{diffGillVm}) can be used to construct computational approaches 
to speed-up simulations of the standard compartment-based model,
because it does not simulate all diffusion events for chemical
species with large diffusion constants~\cite{LiCao2012,LiCao2013}. 
For example, the illustrative simulation in Figure \ref{figure3}
simulates ten times less compartments for $V$ and is less
computationaly intensive than the original simulation in Figure \ref{figure1}.
However, in this work, we are interested in a different question 
than discussing different numerical errors with different discretization 
strategies. We will investigate the Turing pattern formation 
under different discretizations. We will argue that the classical
compartment-based approach is not the best starting point to 
analyse noise in systems which have chemical species with 
different diffusion constants. This conclusion can be already 
demonstrated if we consider a simple two-compartment model as 
we will see in the next section.

\section{Analysis of compartment-based models for $K_u = 2$}
\label{sec3}

We will consider that the domain $[0,L]$ is divided into two compartments
in the $U$ variable, i.e. $K_u = 2.$ Then we have two possible options
for the discretization of the quickly diffusing chemical species $V$:

\medskip

{
\leftskip 10mm

\parindent -5mm

(i) $\gamma = 1$ which corresponds to the classical compartment-based
model where $K_v = 2$;

\parindent -6mm

(ii) $\gamma = 2$ which corresponds to the
generalized compartment-based model where $K_v = 1$. 

}

\medskip

\noindent
We will start with the latter case
which includes three variables $U_1$, $U_2$ and $V_1$ and is easier
to analyse. In Section \ref{seccla2} we compare our results with the
classical compartment-based approach.

\subsection{Generalized compartment-based model: $K_u = 2$ and $K_v = 1$} 

\begin{figure}
\picturesABal{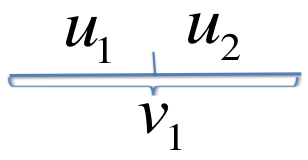}{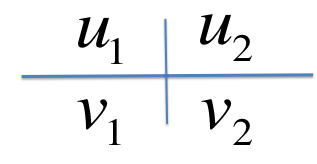}{2cm}{2mm}{12mm}
\caption{
(a) {\it
Generalized compartment-based model for $K_u = 2$ and
$K_v = 1$: The interval is divided into two compartments for $U$ 
and remains as one compartment for $V$.}
(b) {\it
Classical compartment-based model: The interval is divided 
into two compartments for both $U$ and $V$.}}
\label{fig4}
\end{figure}

\begin{figure}
\picturesAB{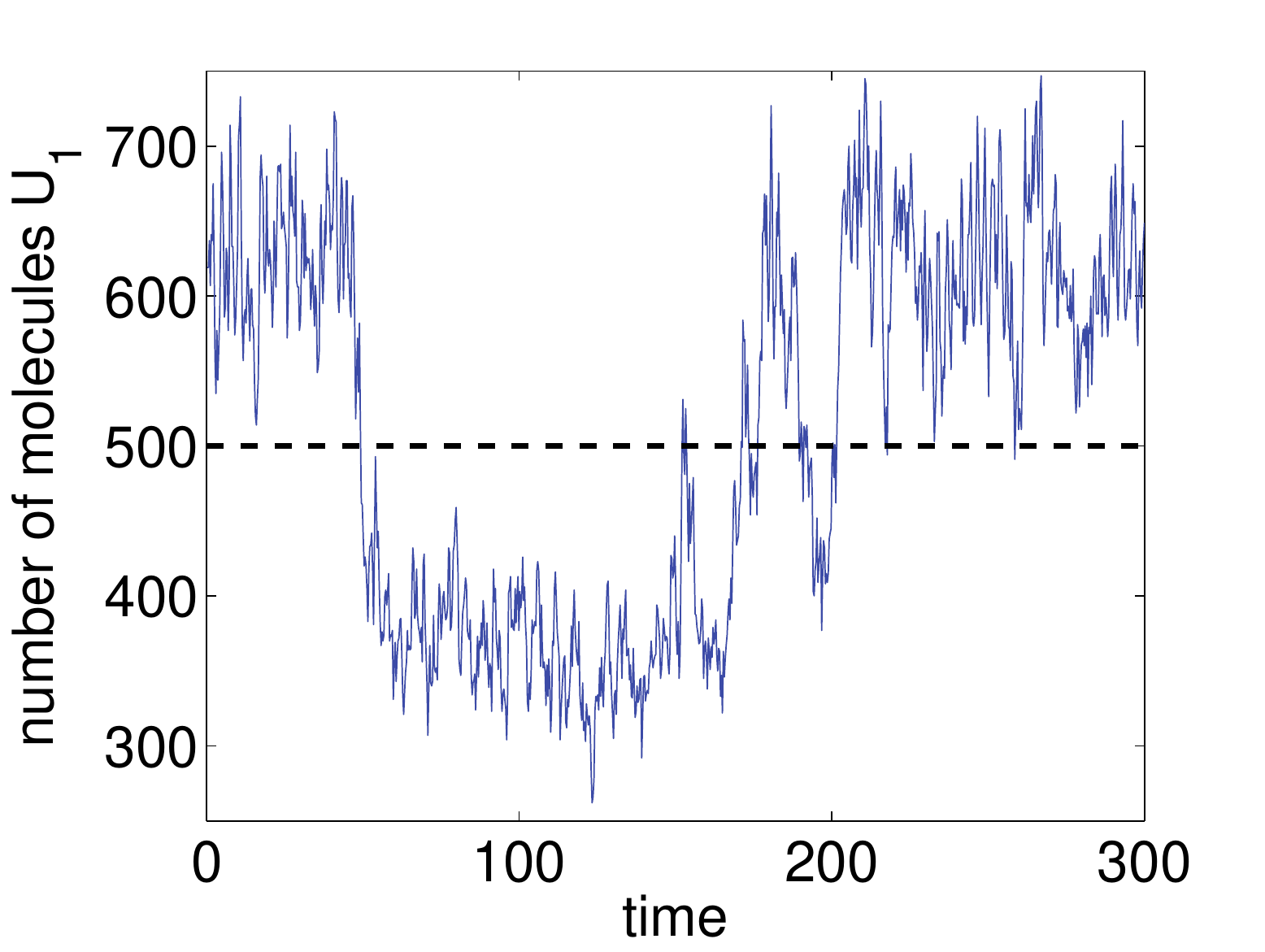}{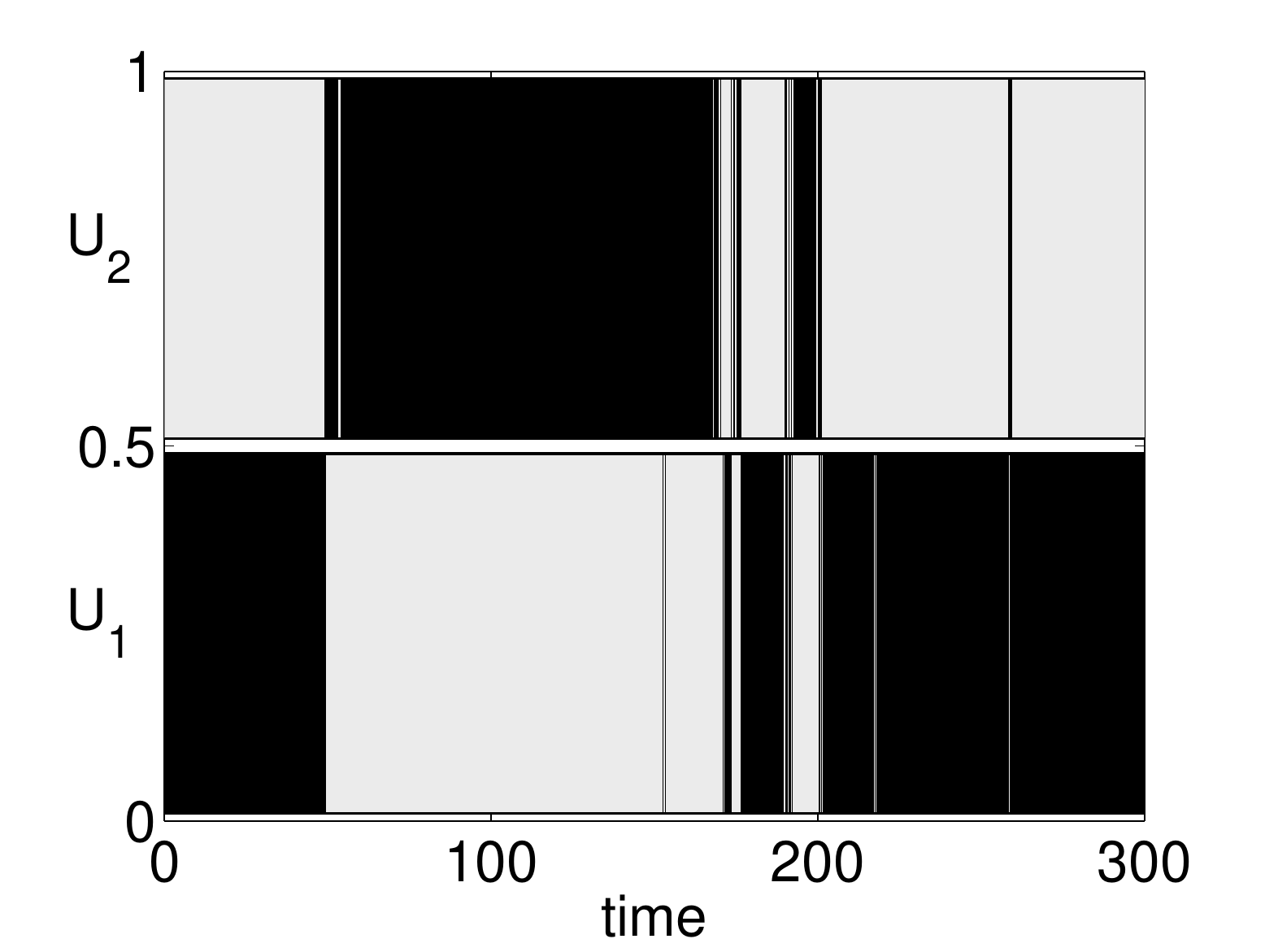}{4.5cm}{5mm}
\caption{{\it {\rm (a)} The time evolution of $U_1$ 
computed for the generalized compartment-based model
with $K_u = 2$ and $K_v = 1$. 
The homogeneous steady state $u_s^2 = 500$ is plotted 
using the dashed line.
{\rm (b)} The time-dependent pattern given by the values
of $U_1$ and $U_2$ computed for the same realization of
the Gillespie algorithm as in the panel {\rm (a)}.}}
\label{figure5}
\end{figure}

We consider the case where the whole interval $[0, \ L]$ 
is divided into two compartments for $U$ and one compartment for $V$. 
The discretization is illustrated in Figure \ref{fig4}(a). 
We will denote by $u_1$, $u_2$ and $v_1$ the average numbers of
molecules of $U_1$, $U_2$ and $V_1$ as predicted by the corresponding 
mean-field model. They satisfy the following system of three ODEs
\cite{Erban:2007:PGS}
\begin{eqnarray} 
\frac{\du_1}{\dt} & =& d_u (u_2 - u_1) + k_1 h_u - k_2 u_1 + 
\frac{k_4}{h_u h_v} u_1^2 v_1, \\
\frac{\du_2}{\dt} & =& d_u (u_1 - u_2) + k_1 h_u - k_2 u_2 + 
\frac{k_4}{h_u h_v} u_2^2 v_1, \\
\frac{\dv_1}{\dt} & = & k_3 h_v 
- \frac{k_4}{h_u h_v} \left(u_1^2 + u_2^2 \right) v_1. 
\end{eqnarray}
We will study the stability of its steady states. 
In order to find the steady state, we let the left hand side terms 
be zero. The corresponding algebraic equations can be written in the
following form: 
\begin{eqnarray} \label{equil1} 
 d_u(u_2 - u_1) + \frac{k_1 L}{2} -  k_2 u_1 + 
 \frac{2 k_4}{L^2} \, u_1^2 v_1 & = & 0, \\
 d_u(u_1 - u_2) + \frac{k_1 L}{2} -  k_2 u_2 
 + \frac{2 k_4}{L^2} \, u_2^2 v_1 & = & 0, \\
 k_3 L - \frac{2 k_4}{L^2} \, (u_1^2 + u_2^2) v_1 & = & 0, 
 \label{equil3} 
\end{eqnarray}
where we used $h_u = L/K_u = L/2$ and $h_v = L/K_v = L.$
Adding all three equations we have 
\begin{equation} \label{sum1} 
	u_1 + u_2 =  \frac{ ( k_1+k_3) L}{ k_2} = 2 \omega L ,  
\end{equation} 
where we used the parameter choice (\ref{omegascaling}).
Let $u_1 = (1 + r) \omega L $ and $u_2 = (1 - r) \omega L$. 
Solving (\ref{equil3}) for $v_1$, we obtain
\begin{equation} \label{v1} 
	v_1 = \frac{k_3L^3}{2k_4 (u_1^2 + u_2^2)} 
	= \frac{3 \omega L}{4 (1 +  r^2 )}.  
\end{equation} 
Substituting \eqref{v1} back to \eqref{equil1}, we have 
$$
	-2 d_u \, r \, \omega L + \frac{k_1 L}{2} - k_2 (1+r) \omega L
	+ 2 k_4 (1 + r)^2 \omega^2 \frac{3 \omega L}{4 (1+r^2)} 
	= 0. 
$$
Using the parameter choice (\ref{omegascaling}), we can simplify it
to
\begin{equation} \label{s1} 
	r \left[ (1- 2d_u) - 2 (1+d_u) r^2 \right] = 0. 
\end{equation} 
The system will have a non-homogeneous solution $u_1 \ne u_2$ if and only
if the equation \eqref{s1} has a non-zero solution, and that requires 
$2 d_u < 1$. Using (\ref{dudvformulam}) and $h_u = L/2$, we obtain
\begin{equation} \label{condition1} 
	D_u < \frac{L^2}{8}.  
\end{equation} 
If this condition is satisfied than the system has two non-nonhomogeneous 
steady-state solutions 
\begin{equation}
u_1 = (1 \pm r) \omega L, \qquad 
u_2 = (1 \mp r) \omega L, \qquad
v_1 = \frac{3 \omega L}{4 (1 +  r^2 )},
\label{solu1u2v1}
\end{equation}
where
\begin{equation}
r
=
\sqrt{\frac{L^2 - 8 D_u}{2 L^2 + 8 D_u}}.
\label{rsol1}
\end{equation} 
In Figure \ref{figure5}, we illustrate this result. We use $L=1$,
$D_u = 0.1$ and $\omega = 500$. Then $r=0.27$ and the steady 
state values of $u_1$ (resp. $u_2$ are): 
$$
u_s^1 \doteq 366, \qquad
u_s^2 \doteq 500, \qquad
u_s^3 \doteq 634.
$$
In Figure \ref{figure5}(a), we present the time evolution of 
$U_1$ computed by the Gillespie algorithm. We initialize the
system at the steady state $[U_1(0),U_2(0),V_1(0)]=[634,366,350]$.
We clearly see that the system is capable of switching between
this state and the second non-homogeneous state. In Figure
\ref{figure5}(b), we visualize the corresponding time-dependent
pattern. As in Figures \ref{figure1} and \ref{figure3}, we plot
the values which are larger than the homogeneous steady state
$u_s^2 = 500$ in black. Light gray colour denotes the values which
are lower than $u_s^2 = 500$. We plot both $U_1$ and $U_2$ values
in Figure \ref{figure5}(b) to visualize the resulting pattern.

\subsection{Classical compartment-based model: $K_u = 2$ and $K_v = 2$} 
\label{seccla2}

\begin{figure}
\centerline{\includegraphics[height=6cm]{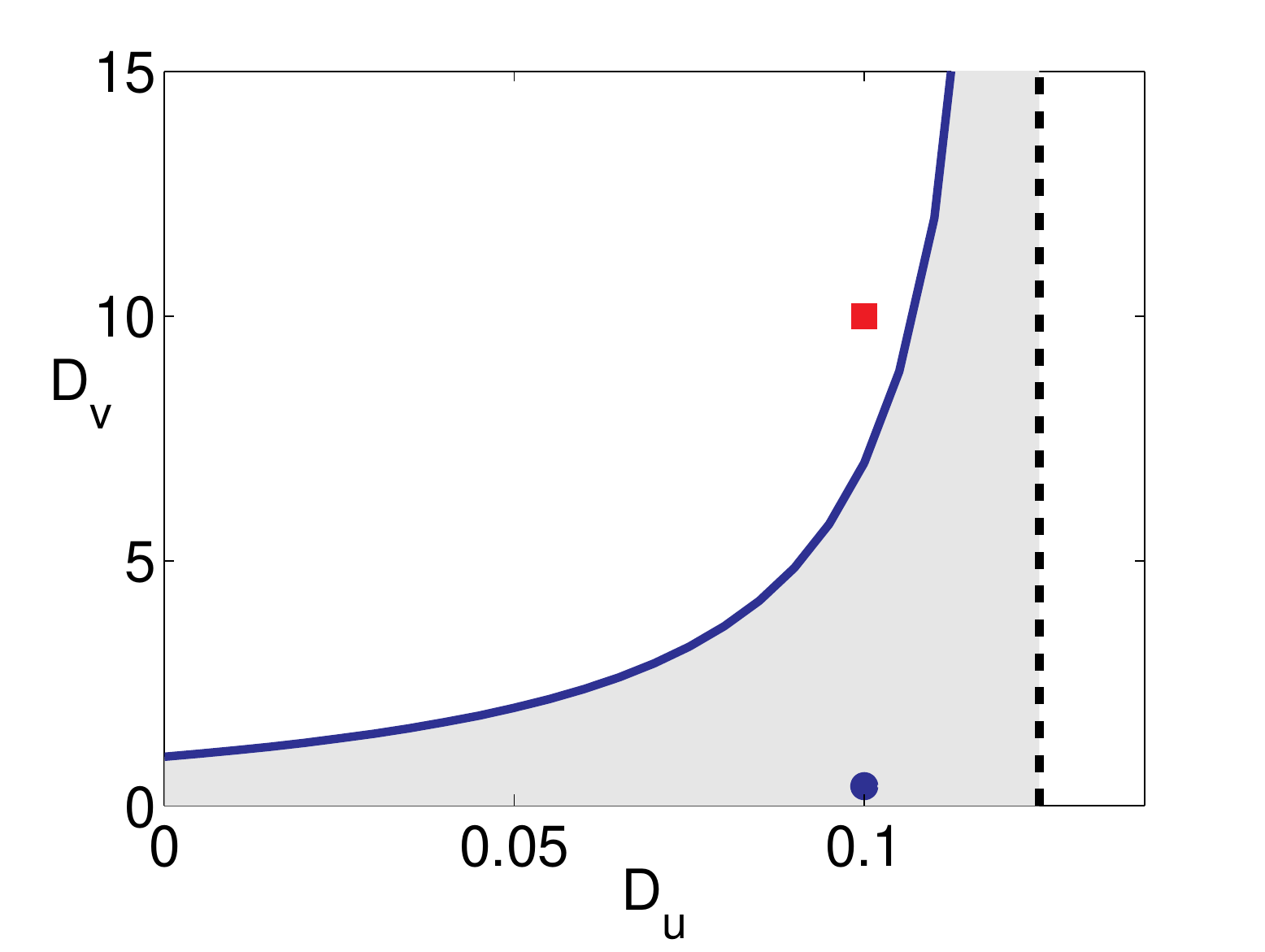}}
\caption{{\it
The regions of patterning in $D_u$-$D_v$ plane. The shaded
area is the region where the standard compartment-based model
does not yield patterns and the generalized compartment-based
model has patterns. The (blue) circle is the parameter regime used
in Figure {\rm \ref{fig7}(a)} and the (red) square is the parameter regime
used in Figure {\rm \ref{fig7}(b)}.}}
\label{fig6}
\end{figure}

\begin{figure}
\picturesAB{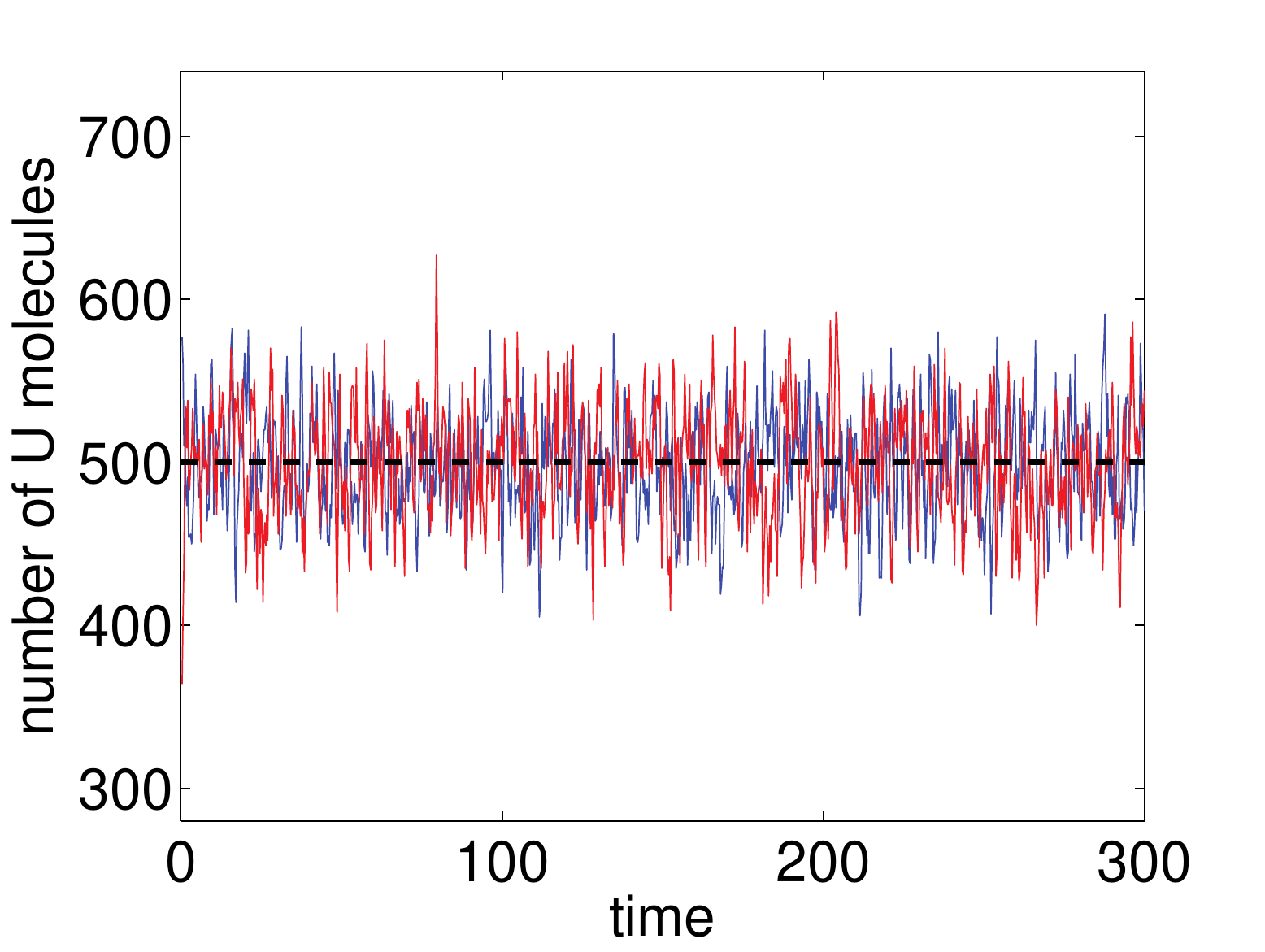}{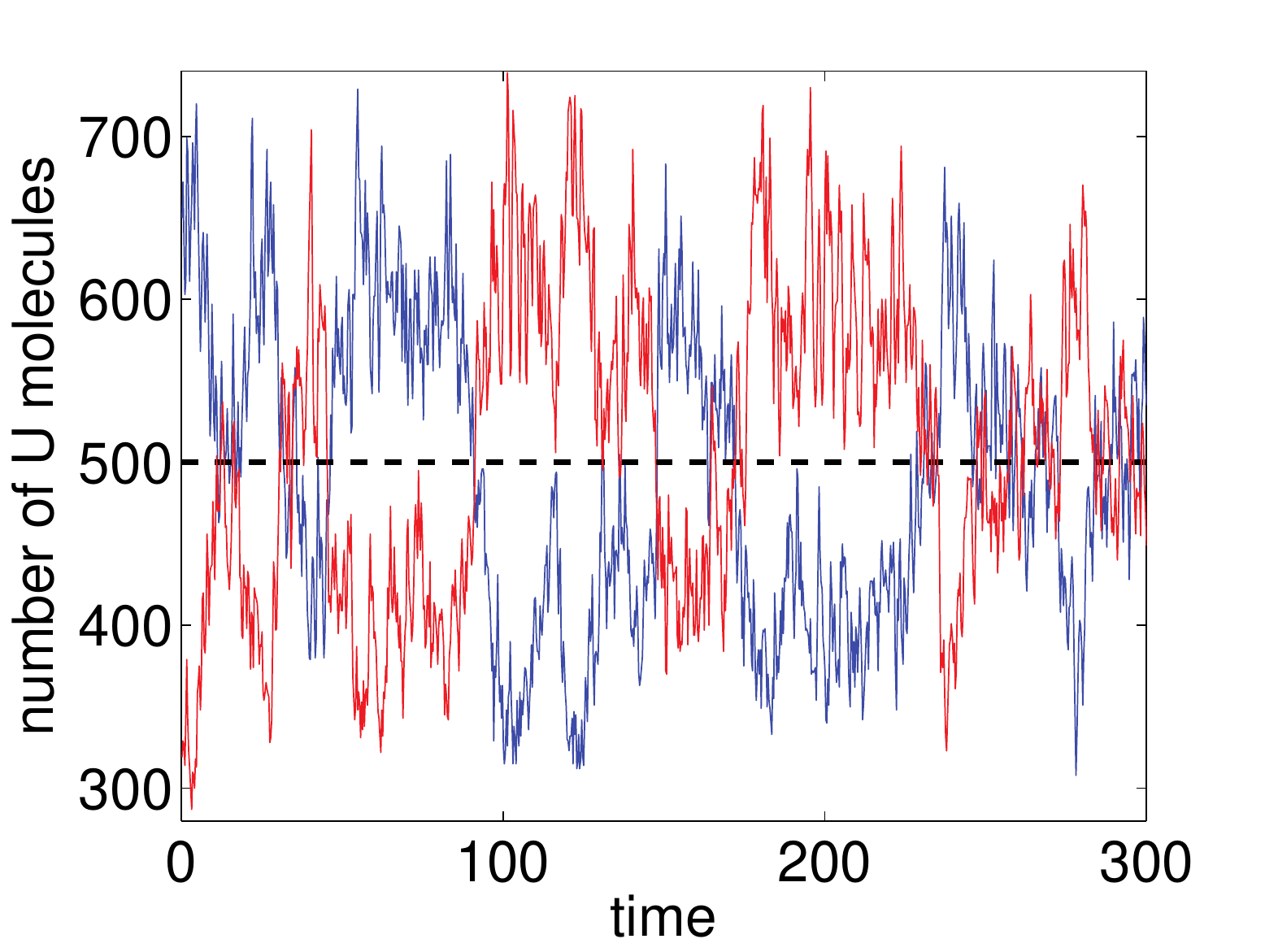}{4.5cm}{5mm}
\picturesnoAB{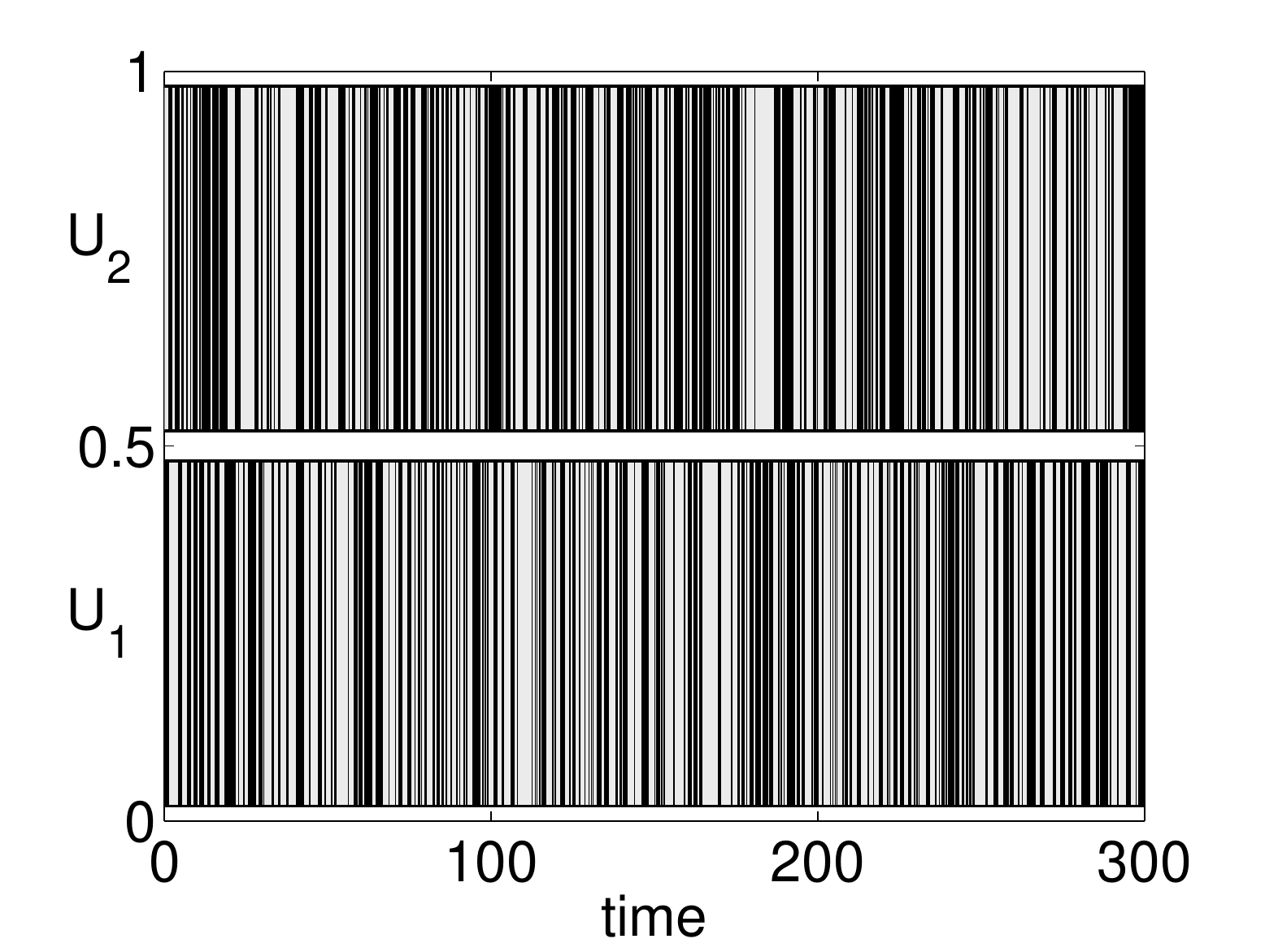}{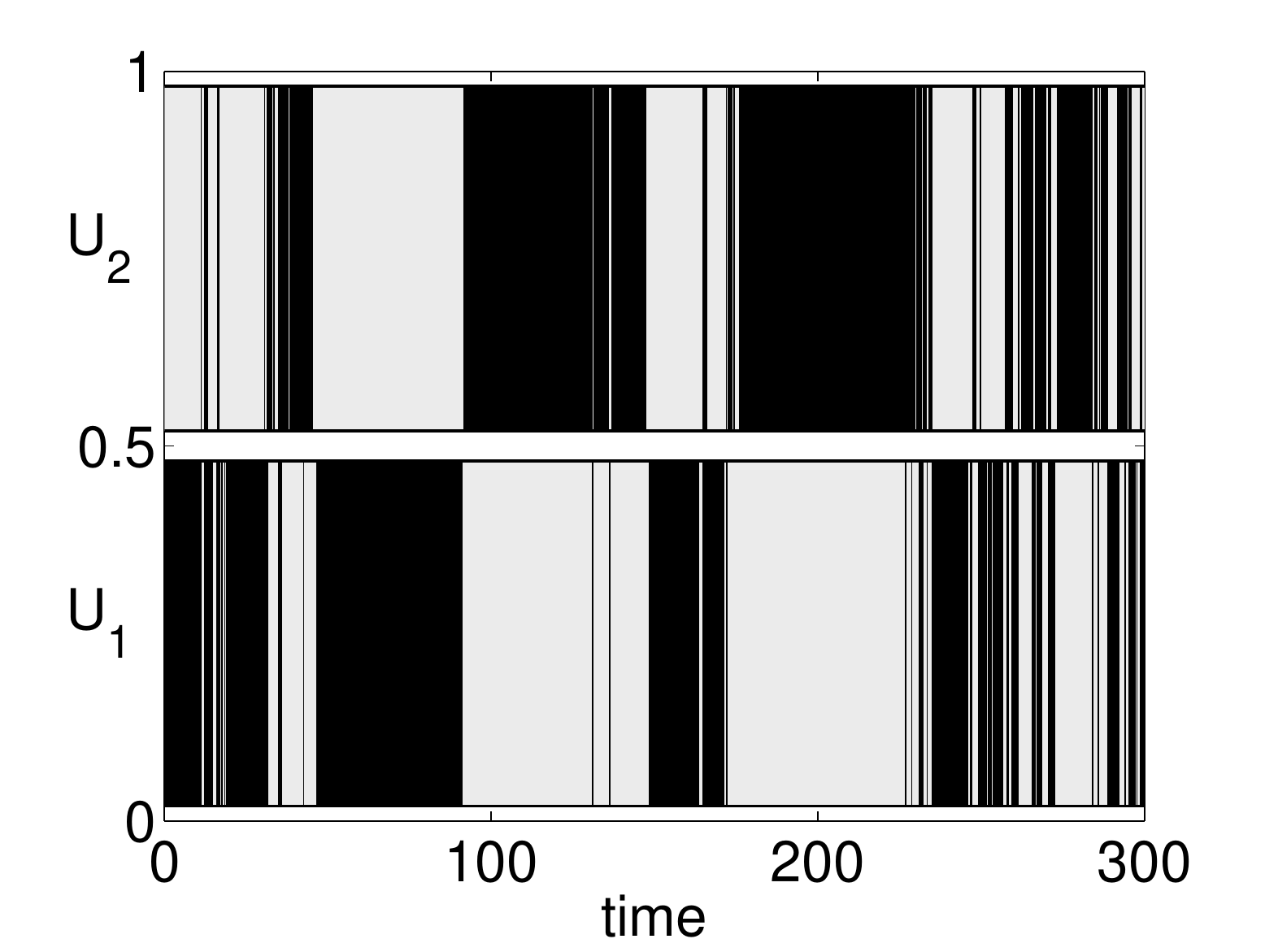}{4.5cm}{5mm}
\caption{{\it Time evolution of $U_1$ (blue line) and $U_2$ (red line) 
for $K_u = K_v = 2$
is shown in top panels for {\rm (a)} $D_u = 0.1$, $D_v = 0.4$
and {\rm (b)}  $D_u = 0.1$, $D_v = 10$. The corresponding
time-dependent pattern is shown in bottom panels.}}
\label{fig7}
\end{figure}

Next we consider the case where the whole interval 
$[0, \ L]$ is divided into two compartments for both $U$ and $V$. 
The discretization is illustrated in 
Figure \ref{fig4}(b). 
Denoting $u_1$, $u_2$, $v_1$ and $v_2$ the average numbers of molecules
obtained by the corresponding mean-field model, they satisfy the
following system of four ODEs
\cite{Erban:2007:PGS}
\begin{eqnarray*} 
\frac{\du_1}{\dt} & =& d_u(u_2 - u_1) + k_1 h_u - k_2 u_1 
+ \frac{k_4}{h_u h_v} u_1^2 v_1, \\
\frac{\du_2}{\dt} & =& d_u(u_1 - u_2) + k_1 h_u - k_2 u_2 
+ \frac{k_4}{h_u h_v}  u_2^2 v_2 , \\
\frac{\dv_1}{\dt} & =& d_v (v_2 - v_1) + k_3 h_v 
- \frac{k_4}{h_u^2} u_1^2 v_1, \\
\frac{\dv_2}{\dt} & =& d_v (v_1 - v_2) + k_3 h_v 
- \frac{k_4}{h_u^2} u_2^2 v_2.  
\end{eqnarray*} 
Again letting the left hand side terms be zero and
using $h_u = h_v = L/2$, we obtain the following system of
algebraic equations
\begin{eqnarray} 
2 d_u(u_2 - u_1) + k_1 L - 2 k_2 u_1 
+ \frac{8 k_4}{L^2} u_1^2 v_1 & = & 0, \label{algebra21}\\
2 d_u(u_1 - u_2) + k_1 L - 2 k_2 u_2 
+ \frac{8 k_4}{L^2}  u_2^2 v_2  & = & 0, \label{algebra22}\\
2 d_v (v_2 - v_1) + k_3 L 
- \frac{8 k_4}{L^2} u_1^2 v_1 & = & 0, \label{algebra23}\\
2 d_v (v_1 - v_2) + k_3 L 
- \frac{8 k_4}{L^2} u_2^2 v_2 & = & 0. \label{algebra24}
\end{eqnarray} 
Adding all equations together, we have 
\begin{equation} \label{sum1m} 
	u_1 + u_2 =  \frac{(k_1+k_3) L}{k_2} = 2 \omega L.  
\end{equation} 
Adding (\ref{algebra23}) and (\ref{algebra24}), we also have 
\begin{equation} \label{v2} 
	u_1^2v_1 + u_2^2v_2 = \frac{k_3 L^3}{4 k_4} 
	= \frac{3 \omega^3 L^3}{4}.  
\end{equation} 
Adding (\ref{algebra21}) and (\ref{algebra23}), we obtain 
\begin{equation} 
\label{pom1}
(k_1 + k_3)L - 2 k_2u_1 + 2d_u(u_2 - u_1) + 2d_v (v_2 - v_1) = 0. 
\end{equation} 
Using (\ref{sum1m}), we have 
$u_1 = (1 + r) \omega L $ and $u_2 = (1 - r) \omega L$ 
for a suitable $r$. Thus (\ref{pom1}) can be rewritten as
\begin{equation} \label{v1v2}
	v_2 - v_1 = \frac{2 r (1+d_u) \omega L}{d_v}
	= 2 r R \omega L, 
\end{equation} 
where we denoted $R = (1+d_u)/d_v$.  
Substituting \eqref{v1v2} into \eqref{algebra23} and denoting
$S = 1 + d_u = d_v R$, we have 
\begin{equation} \label{v1_2}
	v_1 = \frac{(3 + 4Sr)\omega L}{8(1+r)^2}. 
\end{equation} 
Similarly from \eqref{algebra24} we have 
\begin{equation} \label{v2_2}
	v_2 = \frac{(3 - 4Sr)L\omega}{8(1-r)^2}. 
\end{equation} 
Substituting both \eqref{v1_2} and \eqref{v2_2} to \eqref{v1v2}, we obtain 
\begin{equation*}
\frac{3 - 4Sr}{8(1 - r)^2} - \frac{3 + 4Sr}{8(1 + r)^2} = 2Rr. 
\end{equation*} 
which can be simplified to the equation
\begin{equation*} 
	r \left( 4R \left(1 - r^2 \right)^2 
	+ 2 S \left(1 + r^2 \right) - 3 \right) = 0. 
\end{equation*}
We are looking for the non-homogeneous solution where
$r \ne 0$. Denoting $y = r^2 > 0$, we have a quadratic 
equation
\begin{equation} \label{finalequation} 
	4R y^2 + (2S - 8R) y + (4R + 2S - 3) = 0.  
\end{equation}
We will look for conditions such that 
the equation \eqref{finalequation} has a solution $0 < y < 1$ 
(since $-1 < r < 1$). Let
\begin{equation}
	f(y) = 4R y^2 + (2S - 8R) y + (4R + 2S - 3). 
\end{equation} 
Then we have $f(1) = 4S - 3 = 1 + 4d_u > 0$. One can verify that if $f(0) > 0$, it is 
impossible for the equation $f(y) = 0$ to have a solution between 0 and 1. 
On the other hand, if $f(0) < 0$, we will definitely have a solution 
between 0 and 1. Thus we have a necessary and sufficient condition
\begin{equation} 
	f(0) = 4R + 2S - 3 < 0, 
\end{equation} 
which corresponds to the condition for $d_u$ and $d_v$: 
\begin{equation*} 
	\frac{4}{d_v} + 2 < \frac{3}{1 + d_u}. 
\end{equation*} 
We note that $d_u = D_u/h^2$ and $d_v = D_v/h^2$, 
where $h= h_u= h_v = L/2$. Thus the necessary and sufficient
condition for patterns becomes
\begin{equation} \label{condition2_1}
	\frac{L^2}{D_v} + 2 < \frac{3L^2}{L^2 + 4 D_u}. 
\end{equation}
If $D_v \rightarrow \infty$, then the condition (\ref{condition2_1})
becomes the condition \eqref{condition1} which was derived
for the case of the generalized compartment-based model. 
The condition \eqref{condition1} is a necessary condition 
for \eqref{condition2_1} but not sufficient. We illustrate
it in Figure \ref{fig6} for $L=1$. 
The condition \eqref{condition1} corresponds to all parameter
values to the left of the dashed line in Figure \ref{fig6}.
The condition \eqref{condition2_1}
corresponds to the values of $D_u$ and $D_v$ which are above the (blue)
solid line. The shaded area are parameter values for which the
generalized compartment-based model yields non-homogeneous patterns
and the standard compartment-based model does not. Next, we will
use the same value of $D_u$ as in Figure \ref{figure5}, namely
$D_u = 0.1$. We choose two values of $D_v$ which are denoted as
the (blue) circle and (red) square in Figure \ref{fig6}.
We use the Gillespie algorithm to simulate the standard compartment-based
model for $K_u = K_v = 2$. The results are shown in Figure \ref{fig7}.
The top panels show the time evolution of $U_1$ and $U_2$. 
We clearly see the switching between two patterns for $D_v = 10$,
but there is no bistability for $D_v = 0.4.$ The resulting patterns
are visualized in the bottom panels. As in Figures \ref{figure1},
\ref{figure3} and \ref{figure5}, we plot
the values which are larger than the homogeneous steady state
$u_s^2 = 500$ in black. Light gray colour denotes the values which
are lower than $u_s^2 = 500$. 

Let us note that we are comparing the generalized compartment-based
model with $K_u = 2$ and $K_v = 1$ with the classical compartment-based
model. In particular, the generalized compartment-based model
uses $\gamma = 2.$ If we substitute $\gamma = 2$ in 
formula (\ref{gammaformula}),
we obtain $D_v = 4 D_u$. In particular, the parameter values 
$D_u = 0.1$ and $D_v = 0.4.$ are compatible with the choice
(\ref{gammaformula}). However, the standard comparment-based
model does not exhibit patterns for this parameter choice as we
observed in Figure \ref{fig7}(a). 

\medskip

\noindent
{\bf Remark.} Let $z = L^2$. Then the inequality (\ref{condition2_1})
becomes 
\begin{equation} 
	z^2 + (4D_u - D_v) z + 8 D_uD_v < 0, 
\end{equation} 
which is possible for some values of $L$ if and only if 
\begin{equation} 
	4D_u < D_v \quad \mbox{and} \quad (4D_u - D_v)^2 - 32D_uD_v > 0. 
\end{equation}
Thus patterns are possible for some values of $L$ provided that 
\begin{equation}  \label{Turingcondition}
	\frac{D_v}{D_u} > 20 + 8 \sqrt{6} \approx 39.6 . 
\end{equation} 
This condition is also 
the condition for the Turing patterns to show for the 
original system of mean-field partial differential equations 
(\ref{uPDEschnak})--(\ref{vPDEschnak}).

\section{Comparison of compartment-based models for $K_u > 2$} 
\label{sec4}

\begin{figure}
\picturesAB{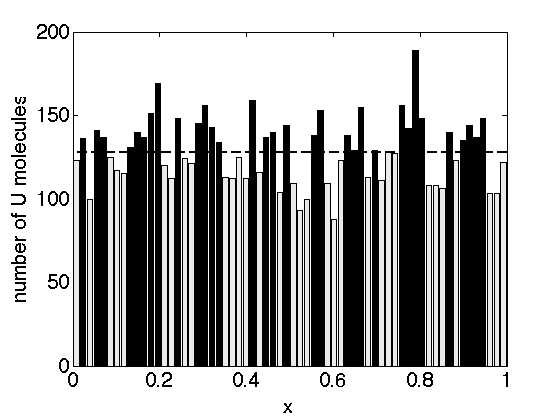}{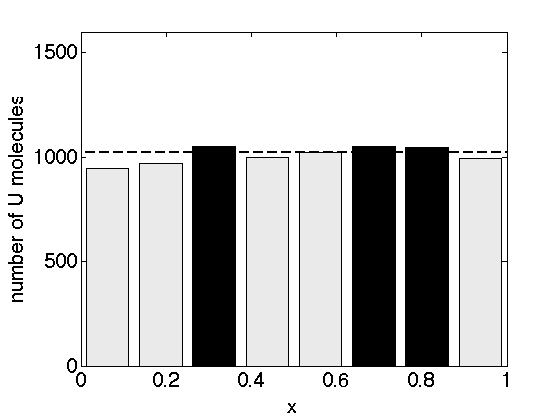}{4.5cm}{5mm}
\picturesCD{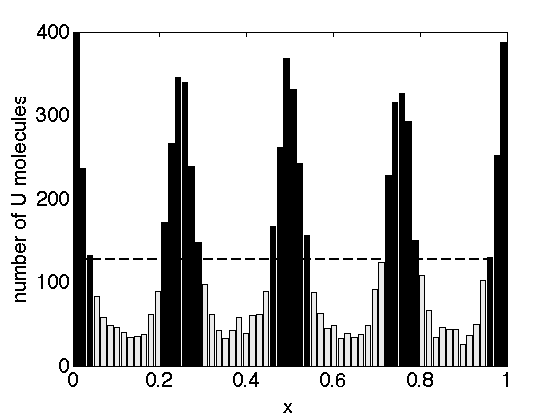}{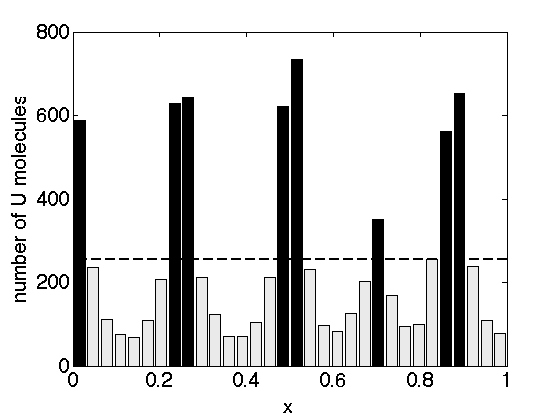}{4.5cm}{5mm}
\caption{{\it Spatial distribution of $U$ at time $T = 100$ 
for $D_v = 20 D_u$, $\omega=4096$ and $D_u = 5\times 10^{-4}$
with}
(a) 
{\it $K_u = K_v = 64$;}
(b) 
{\it $K_u = K_v = 8$;} 
(c) 
{\it $K_u = 64$ and $K_v = 8$;}
(d) 
{\it $K_u = 32$ and $K_v = 8$. 
There is no Turing pattern in the top panels (classical 
compartment-based model). Turing patterns appear in the
bottom panels (generalized compartment-based model).}
}
\label{fig8}
\end{figure}

The condition \eqref{condition1} for the generalized compartment-based
model is only a necessary condition for the condition \eqref{condition2_1} 
for the classical case as we showed in Figure \ref{fig6}. 
The bistability condition difference suggests that, if we use different 
discretizations for $U$ and $V$, the stability of the homogeneous
system may change. In this section, we compare the generalized and
classical compartment-based models for $K_u > 2.$ In Figure \ref{fig8},
we use $D_u = 5\times 10^{-4}$ and $D_v = 20 D_u$. In this case the
condition for (deterministic) Turing patterns \eqref{Turingcondition}
is not satisfied. The classical compartment-based model also 
does not show Turing patterns as it is demonstrated in 
Figure \ref{fig8}(a) (with $K_u = K_v = 64$ compartments) and Figure 
\ref{fig8}(b) (with $K_u = K_v = 8$ compartments). 
In both cases, no spatial Turing 
pattern is observed except noise from stochastic effect. 
However, if the generalized compartment-based model is used, 
then the Turing pattern may appear. In Figure \ref{fig8}(c),
a result for the generalized compartment-based model 
with $K_u = 64$ and $K_v = 8$ is presented.
There is a clear Turing pattern. 
In Figure \ref{fig8}(c), we have $\gamma = 8$. We also tested cases when  
$\gamma = 2$ and $\gamma = 4$ and obtained Turing patterns. 
The case $\gamma = 4$ is plotted in Figure \ref{fig8}(d).

\begin{figure}
\picturesAB{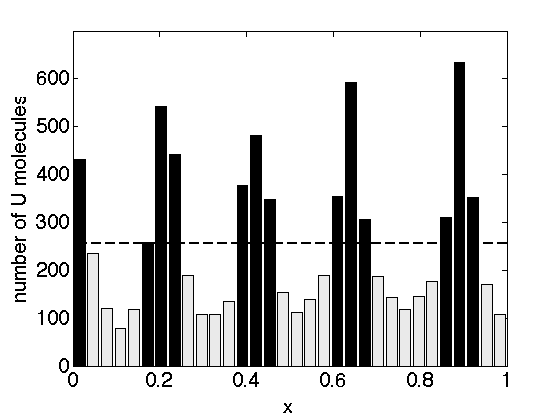}{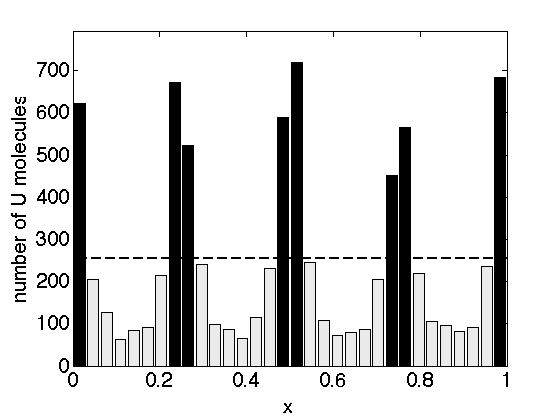}{4.5cm}{5mm}
\caption{
{\it Spatial distribution of $U$ at time $T = 100$ 
for $D_v = 80 D_u$. Both discretization strategies clearly 
show Turing patterns. We use $\omega=4096$, $D_u = 5\times 10^{-4}$
with}
(a) {\it $K_u = K_v = 32$;}
(b) {\it $K_u = 32$ and $K_v = 8$.}}
\label{fig9}
\end{figure}

\begin{figure}
\picturesAB{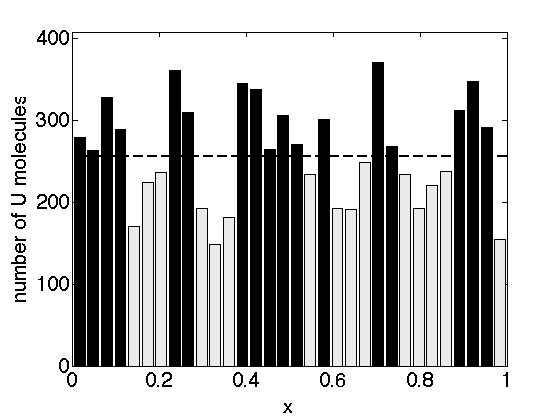}{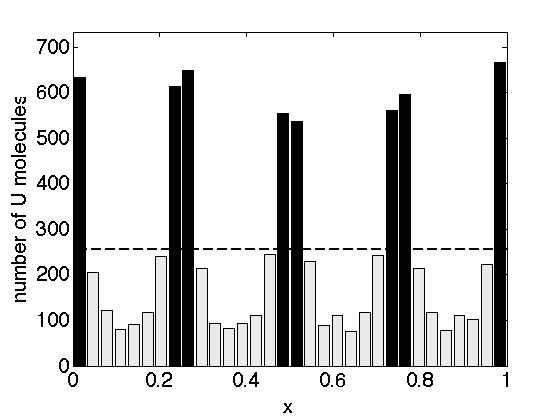}{4.5cm}{5mm}
\picturesCD{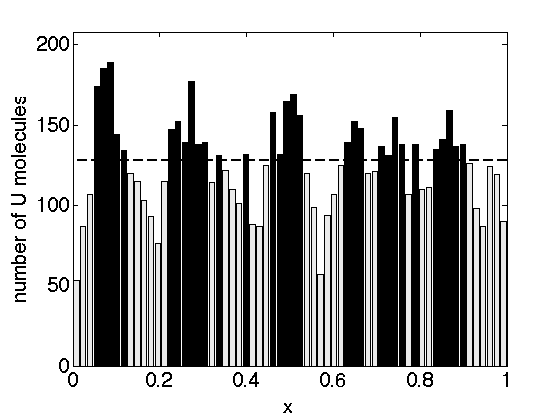}{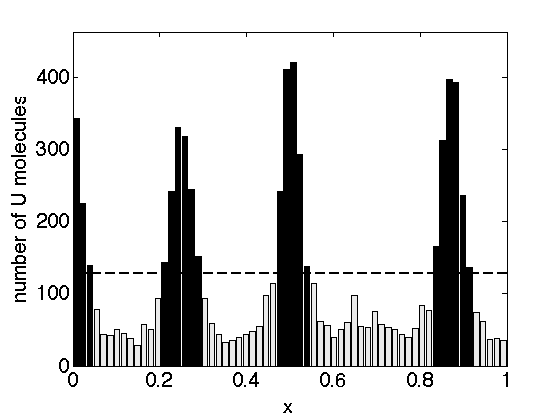}{4.5cm}{5mm}
\caption{
{\it Spatial distribution of $U$ at time $T = 100$ 
for $D_v = 40 D_u$. The generalized compartment-based
model clearly shows Turing patterns, while it is difficult 
to see whether Turing patterns appear in the classical
compartment-based model. We use $\omega=4096$, 
$D_u = 5\times 10^{-4}$ with}
(a) {\it $K_u = K_v = 32$;}
(b) {\it $K_u = 32$ and $K_v = 8$;}
(c) {\it $K_u = K_v = 64$;}
(d) {\it $K_u = 64$ and $K_v = 8$.}}
\label{fig10}
\end{figure}

In Figure \ref{fig9}, we demonstrate that both discretizations
strategies clearly show Turing patterns when we increase
the ratio of diffusion constants to $D_v/D_u = 80$. 
In this case, the condition for (deterministic) Turing 
patterns \eqref{Turingcondition} is satisfied. Finally,
we present results for $D_v = 40D_u$ in Figure \ref{fig10}. 
In the deterministic PDE system, when $D_v = 40D_u$, Turing pattern 
should still appear. But in the classical compartment-based
model, it is hard to claim that there is a visible Turing pattern 
(see Figures \ref{fig10}(a) and \ref{fig10}(c)).
Considering the generalized compartment-based model,
Turing patterns can be clearly observed
(see Figures \ref{fig10}(b) and \ref{fig10}(d)).  

\section{Discussion}
\label{sec5}

We showed that two choices of compartments illustrated in Figure
\ref{figure2} can give different parameter regions for stochastic
Turing patterns. An obvious question is which one is correct.
One possibility to address this question is to consider a more
detailed molecular-based approach which would be written in the
form of Brownian dynamics \cite{Erban:2009:SMR}. We are currently
working on such a simulation and we will report our findings in
a future publication.

Although our results might look like a warning against the use of
compart\-ment-based methods for patterns based on the Turing mechanism, 
there are very good reasons to use the compart\-ment-based model
in other situations \cite{Engblom:2009:SSR,Isaacson:2006:IDC}.
Compartment-based models are often less computationally intensive
than detailed Brownian dynamics simulations
\cite{Flegg:2012:TRM,Hellander:2012:CMM}. They can be used
for developing efficient multiscale methods where parts of 
the domain are simulated using the detailed Brownian dynamics
while the rest of the domain is simulated using compartments
\cite{Erban:2013:MRS,Flegg:2013:DSN}. They can also be used
to bridge Brownian dynamics simulations with macroscopic
PDEs \cite{Ferm:2010:AAS}, because direct multiscale methods for
coupling Brownian dynamics with PDEs are challenging to
implement \cite{Franz:2013:MRA}.

We showed in Figure \ref{fig9} that the resulting
patterns are comparable when the ratio
of diffusion constants is sufficiently large. In this case, 
the generalized compartment-based model could also be used to construct 
computational approaches to speed-up simulations of the standard 
compartment-based model, because it does not simulate all diffusion 
events for chemical species with large diffusion 
constants~\cite{LiCao2012,LiCao2013}. 

\section*{Acknowledgements}

The research leading to these results has received funding from
the European Research Council under the {\it European Community's}
Seventh Framework Programme {\it (FP7/2007-2013)} /
ERC {\it grant agreement} No. 239870. 
This publication
was based on work supported in part by Award No KUK-C1-013-04, made by King
Abdullah University of Science and Technology (KAUST).
Radek Erban would also like to thank the Royal Society for
a University Research Fellowship; Brasenose College, University of Oxford,
for a Nicholas Kurti Junior Fellowship; and the Leverhulme Trust for
a Philip Leverhulme Prize. 
Yang Cao's work was supported by the National Science
Foundation under awards DMS-1225160 and CCF-0953590,
and the National Institutes of Health under award GM078989.


\begin{thebibliography}{10}

\bibitem{Asslani:2012:STP}
M.~Asslani, F.~Di~Patti, and D.~Fanelli.
\newblock Stochastic {T}uring patterns on a network.
\newblock {\em Physical Review E}, 86:046105, 2012.

\bibitem{Barrass:2006:MTM}
I.~Barrass, E.~Crampin, and P.~Maini.
\newblock Mode transitions in a model reaction-diffusion system driven by
  domain growth and noise.
\newblock {\em Bulletin of Mathematical Biology}, 68:981--995, 2006.

\bibitem{Biancalani:2010:STP}
T.~Biancalani, D.~Fanelli, and F.~Di~Patti.
\newblock Stochastic {T}uring patterns in a {B}russelator model.
\newblock {\em Physical Review E}, 81:046215, 2010.

\bibitem{Biancalani:2011:SWB}
T.~Biancalani, T.~Galla, and A.~McKane.
\newblock Stochastic waves in a {B}russelator model with nonlocal interaction.
\newblock {\em Physical Review E}, 84:026201, 2011.

\bibitem{Black:2012:SFE}
A.~Black and A.~McKane.
\newblock Stochastic formulations of ecological models and their applications.
\newblock {\em Trends in Ecology and Evolution}, 27(6):337--345, 2012.

\bibitem{Butler:2011:FTP}
T.~Butler and N.~Goldenfeld.
\newblock Fluctuation-driven {T}uring patterns.
\newblock {\em Physical Review E}, 84:011112, 2011.

\bibitem{Cao:2004:EFS}
Y.~Cao, H.~Li, and L.~Petzold.
\newblock Efficient formulation of the stochastic simulation algorithm for
  chemically reacting systems.
\newblock {\em Journal of Chemical Physics}, 121(9):4059--4067, 2004.

\bibitem{Crampin:1999:RDG}
E.~Crampin, E.~Gaffney, and P.~Maini.
\newblock Reaction and diffusion on growing domains: Scenarios for robust
  pattern formation.
\newblock {\em Bulletin of Mathematical Biology}, 61:1093--1120, 1999.

\bibitem{Engblom:2009:SSR}
S.~Engblom, L.~Ferm, A.~Hellander, and P.~L\"otstedt.
\newblock Simulation of stochastic reaction-diffusion processes on unstructured
  meshes.
\newblock {\em SIAM Journal on Scientific Computing}, 31:1774--1797, 2009.

\bibitem{Erban:2007:RBC}
R.~Erban and S.~J. Chapman.
\newblock Reactive boundary conditions for stochastic simulations of
  reaction-diffusion processes.
\newblock {\em Physical Biology}, 4(1):16--28, 2007.

\bibitem{Erban:2009:SMR}
R.~Erban and S.~J. Chapman.
\newblock Stochastic modelling of reaction-diffusion processes: algorithms for
  bimolecular reactions.
\newblock {\em Physical Biology}, 6(4):046001, 2009.

\bibitem{Erban:2007:PGS}
R.~Erban, S.~J. Chapman, and P.~Maini.
\newblock A practical guide to stochastic simulations of reaction-diffusion
  processes.
\newblock 35 pages, available as http://arxiv.org/abs/0704.1908, 2007.

\bibitem{Erban:2013:MRS}
R.~Erban, M.~Flegg, and G.~Papoian.
\newblock Multiscale stochastic reaction-diffusion modelling: application to
  actin dynamics in filopodia.
\newblock {\em Bulletin of Mathematical Biology}, to appear:DOI:
  10.1007/s11538--013--9844--3, 2013.

\bibitem{Fange:2006:NMP}
D.~Fange and J.~Elf.
\newblock Noise-induced {M}in phenotypes in {E}. coli.
\newblock {\em PLoS Computational Biology}, 2(6):637--648, 2006.

\bibitem{Ferm:2010:AAS}
L.~Ferm, A.~Hellander, and P.~L\"otstedt.
\newblock An adaptive algorithm for simulation of stochastic reaction-diffusion
  processes.
\newblock {\em Journal of Computational Physics}, 229:343--360, 2010.

\bibitem{Flegg:2012:TRM}
M.~Flegg, J.~Chapman, and R.~Erban.
\newblock The two-regime method for optimizing stochastic reaction-diffusion
  simulations.
\newblock {\em Journal of the Royal Society Interface}, 9(70):859--868, 2012.

\bibitem{Flegg:2013:DSN}
M.~Flegg, S.~R\"udiger, and R.~Erban.
\newblock Diffusive spatio-temporal noise in a first-passage time model for
  intracellular calcium release.
\newblock {\em Journal of Chemical Physics}, 138:154103, 2013.

\bibitem{Franz:2013:MRA}
B.~Franz, M.~Flegg, J.~Chapman, and R.~Erban.
\newblock Multiscale reaction-diffusion algorithms: {PDE}-assisted {B}rownian
  dynamics.
\newblock {\em SIAM Journal on Applied Mathematics}, 73(3):1224--1247, 2013.

\bibitem{Fu:2008:SST}
Z.~Fu, X.~Xu, H.~Wang, and Q.~Quoyang.
\newblock Stochastic simulation of {T}uring patterns.
\newblock {\em Chinese Physical Letters}, 25(4):1220--1223, 2008.

\bibitem{Gibson:2000:EES}
M.~Gibson and J.~Bruck.
\newblock Efficient exact stochastic simulation of chemical systems with many
  species and many channels.
\newblock {\em Journal of Physical Chemistry A}, 104:1876--1889, 2000.

\bibitem{Gillespie:1977:ESS}
D.~Gillespie.
\newblock Exact stochastic simulation of coupled chemical reactions.
\newblock {\em Journal of Physical Chemistry}, 81(25):2340--2361, 1977.

\bibitem{Hattne:2005:SRD}
J.~Hattne, D.~Fange, and J.~Elf.
\newblock Stochastic reaction-diffusion simulation with {M}eso{RD}.
\newblock {\em Bioinfor\-matics}, 21(12):2923--2924, 2005.

\bibitem{Hellander:2012:CMM}
A.~Hellander, S.~Hellander, and P.~L\"otstedt.
\newblock Coupled mesoscopic and microscopic simulation of stochastic
  reaction-diffusion processes in mixed dimensions.
\newblock {\em Multiscale Modeling and Simulation}, 10(2):585--611, 2012.

\bibitem{Hellander:2012:RME}
S.~Hellander, A.~Hellander, and L.~Petzold.
\newblock Reaction-diffusion master equation in the microscopic limit.
\newblock {\em Physical Review E}, 85:042901, 2012.

\bibitem{Hori:2012:NSP}
Y.~Hori and S.~Hara.
\newblock Noise-induced spatial pattern formation in stochastic
  reaction-diffusion systems.
\newblock {\em Proc. of 51st IEEE Conference on Decision and Control}, pages
  1053--1058, 2012.

\bibitem{Hu:2013:SAR}
J.~Hu, H.~Kang, and H.~Othmer.
\newblock Stochastic analysis of reaction-diffusion processes.
\newblock {\em Bulletin of Mathematical Biology}, to appear:DOI:
  10.1007/s11538--013--9849--y, 2013.

\bibitem{Isaacson:2009:RME}
S.~Isaacson.
\newblock The reaction-diffusion master equation as an asymptotic approximation
  of diffusion to a small target.
\newblock {\em SIAM Journal on Applied Mathematics}, 70(1):77--111, 2009.

\bibitem{Isaacson:2006:IDC}
S.~Isaacson and C.~Peskin.
\newblock Incorporating diffusion in complex geometries into stochastic
  chemical kinetics simulations.
\newblock {\em SIAM Journal on Scientific Computing}, 28(1):47--74, 2006.

\bibitem{Kang:2012:NMC}
H.~Kang, L.~Zheng, and H.~Othmer.
\newblock A new method for choosing the computational cell in stochastic
  reaction-diffusion systems.
\newblock {\em Journal of Mathematical Biology}, 65(6-7):1017--1099, 2012.

\bibitem{Kepper:1991:TCP}
P.~Kepper, V.~Castets, E.~Dulos, and J.~Boissonade.
\newblock Turing-type chemical patterns in the chlorite-iodide-malonic acid
  reaction.
\newblock {\em Physica D}, 49:161--169, 1991.

\bibitem{LiCao2012}
F.~Li and Y.~Cao. 
\newblock Multiscale discretization for reaction diffusion systems. 
\newblock Proceedings of the 2012 International Conference on Bioinformatics and Computational
Biology (Editors: Hamid R. Arabnia, Quoc-Nam Tran Associate Editors: Andy Marsh, Ashu
M. G. Solo), Las Vegas, Nevada, USA, 2012, 305-311. 

\bibitem{LiCao2013}
F.~Li and Y.~Cao. 
\newblock Optimal discretization size and multigrid discretization
method for 1D multiscale reaction diffusion systems. 
\newblock submitted. 

\bibitem{Maini:2012:TMB}
P.~Maini, T.~Woolley, R.~Baker, E.~Gaffney, and S.~Seirin~Lee.
\newblock Turing's model for biological pattern formation and the robustness
  problem.
\newblock {\em Interface focus}, 2:487--496, 2012.

\bibitem{McKane:2013:SPF}
A.~McKane, T.~Biancalani, and T.~Rogers.
\newblock Stochastic pattern formation and spontaneous polarization: the linear
  noise approximation and beyond.
\newblock {\em Bulletin of Mathematical Biology}, to appear:DOI:
  10.1007/s11538--013--9827--4, 2013.

\bibitem{Murray:2002:MB}
J.~Murray.
\newblock {\em {M}athematical {B}iology}.
\newblock Springer Verlag, 2002.

\bibitem{Qiao:2006:SDS}
L.~Qiao, R.~Erban, C.~Kelley, and I.~Kevrekidis.
\newblock Spatially distributed stochastic systems: Equation-free and
  equation-assisted preconditioned computation.
\newblock {\em Journal of Chemical Physics}, 125:204108, 2006.

\bibitem{Quyang:1991:TUS}
Q.~Quyang and H.~Swinney.
\newblock Transition from a uniform state to hexagonal and striped {T}uring
  patterns.
\newblock {\em Nature}, 352:610--612, 1991.

\bibitem{Satnoianu:2000:TIG}
R.~Satnoianu, M.~Menzinger, and P.~Maini.
\newblock Turing instabilities in general systems.
\newblock {\em J. Math. Biol.}, 41:493--512, 2000.

\bibitem{Schnakenberg:1979:SCR}
J.~Schnakenberg.
\newblock Simple chemical reaction systems with limit cycle behaviour.
\newblock {\em Journal of Theoretical Biology}, 81:389--400, 1979.

\bibitem{Scott:2011:AIN}
M.~Scott, F.~Poulin, and H.~Tang.
\newblock Approximating intrinsic noise in continuous multispecies models.
\newblock {\em Proceedings of the Royal Society A}, 467:718--737, 2011.

\bibitem{Sick:2006:WDD}
S.~Sick, S.~Reinker, J.~Timmer, and T.~Schlake.
\newblock {WNT} and {DKK} determine hair follicle spacing through a
  reaction-diffusion mechanism.
\newblock {\em Science}, 314:1447--1450, 2006.

\bibitem{Turing:1952:CBM}
A.~Turing.
\newblock The chemical basis of morphogenesis.
\newblock {\em Phil. Trans. Roy. Soc. Lond.}, 237:37--72, 1952.

\bibitem{Twomey:2007:SMR}
A.~Twomey.
\newblock On the stochastic modelling of reaction-diffusion processes.
\newblock {M.S}c. {T}hesis, University of Oxford, United Kingdom, September
  2007.

\bibitem{Vigelius:2012:SSP}
M.~Vigelius and B.~Meyer.
\newblock Stochastic simulations of pattern formation in excitable media.
\newblock {\em PLoS ONE}, 7(8):e45208, 2012.

\bibitem{Woolley:2011:PSM}
T.~Woolley, R.~Baker, E.~Gaffney, and P.~Maini.
\newblock Power spectra methods for a stochastic description of diffusion on
  deterministically growing domains.
\newblock {\em Physical Review E}, 84:021915, 2011.

\bibitem{Woolley:2011:SRD}
T.~Woolley, R.~Baker, E.~Gaffney, and P.~Maini.
\newblock Stochastic reaction and diffusion on growing domains: understanding
  the breakdown of robust pattern formation.
\newblock {\em Physical Review E}, 84:046216, 2011.

\bibitem{Woolley:2012:EIS}
T.~Woolley, R.~Baker, E.~Gaffney, P.~Maini, and S.~Seirin-Lee.
\newblock Effects of intrinsic stochasticity on delayed reaction-diffusion
  patterning systems.
\newblock {\em Physical Review E}, 85:051914, 2012.

\end{thebibliography}
\end{document}